\documentclass[aps,prb,showpacs,twoside,amsmath, amsfonts]{revtex4}
\usepackage{psfrag}
\usepackage{graphicx}
\usepackage{epsfig}
\usepackage{amssymb}

\begin{document}
\title{ Random Matrix Theory: Wigner-Dyson statistics and beyond.
\\ Lecture notes given at SISSA (Trieste, Italy)}
\author{V.E.Kravtsov}
\affiliation{The Abdus Salam International Centre for Theoretical
Physics, P.O.B. 586, 34100 Trieste, Italy, \\ Landau Institute for
Theoretical Physics, 2 Kosygina st., 117940 Moscow, Russia.}
\date{\today}
\begin{abstract}

\end{abstract}
 \maketitle
\section{Invariant and Non-invariant Gaussian random matrix ensembles}
Any random matrix ensemble (RME) is determined through the
probability distribution function (PDF) $P({\bf M})$ that depends on
the matrix entries $M_{nm}$. An important special class of random
matrix ensembles is given by the PDF which is invariant under
rotation of basis ${\bf M}\rightarrow {\bf T}{\bf M}{\bf T}^{-1}$:
\begin{equation}
\label{inv} P({\bf M})\propto {\rm exp}\left[ -{\rm Tr}\,V({\bf
M})\right],
\end{equation}
where $V({\bf M})$ is and arbitrary function of ${\bf M}$ analytic
at ${\bf M}=0$. The invariance is ensured by the trace ${\rm Tr}$ in
front of the matrix function $V({\b M})$. The RME with the PDF of
the form Eq.(\ref{inv}) will be referred to as {\it invariant RME}.

In general the PDF Eq.(\ref{inv}) corresponds to a non-trivial
correlation between fluctuating matrix entries. However, there is
one extremely important case when \\
(i) the matrix ${\bf M}$ is Hermitean  ${\bf M}\equiv{\bf H}={\bf
H}^{\dagger}$ and \\
(ii) $V({\bf H})=a{\bf H}^{2}$.\\
 In this case
$${\rm exp}\left[-{\rm Tr V({\bf H})}\right]={\rm exp}\left[- a\sum_{n,m} |H_{nm}|^{2}\right]=\prod_{n,m}{\rm
exp}\left[ -a\,|H_{nm}|^{2}\right]$$ so that all matrix entries
fluctuate independently around zero. This is the celebrated Gaussian
random matrix ensemble of Wigner and Dyson (WD).

Note that Gaussian random matrix ensembles can be also {\it
non-invariant}. The generic non-invariant Gaussian RME is determined
by the PDF of the form:
\begin{equation}
\label{NIE} P({\bf H})\propto {\rm
exp}\left[-\sum_{n,m}\frac{|H_{nm}|^{2}}{A_{nm}} \right].
\end{equation}
In this case each matrix entry fluctuate independently of the other
but with the variance which depends on the indices $n,m$ that label
the matrix entry. The simplest Gaussian non-invariant ensemble is
the {\it Rosenzweig-Porter ensemble} \cite{RosPort} for which
\begin{equation}
\label{RP} A_{nm}=\left\{ \begin{matrix}a, &  n\neq m\cr \Lambda\,a
& n=m\cr
\end{matrix}\right.
\end{equation}
It is remarkable that both the classic WD ensemble and the
Rosenzweig-Porter ensemble   allow for an exact solution
\cite{Mehta, Kunz}.

Physically, an invariant random matrix ensemble describes extended
(but phase-randomized) states, where the localization effects are
negligible. In contrast to that any non-invariant ensemble accounts
for a sort of structure of eigenfunctions (e.g. localization) in a
given basis which may be not the case in a different rotated basis
(remember about the extended states in the tight-binding model which
are the linear combinations of states localized at a given site).

In particular the problem of localization in a quasi-1 wire can
mapped onto the non-invariant {\it banded RME} with the variance
matrix equal to:
\begin{equation}
\label{band} A_{nm}={\rm exp}\left[-|n-m|/B\right]
\end{equation}
This model can be efficiently mapped onto a nonlinear supersymmetric
sigma model and solved by the transfer matrix method
\cite{Mirlin-Fyod}.

Finally, we mention a {\it critical power-law banded random matrix
ensemble} (CPLB-RME) for which the variance matrix is of the
Lorenzian form:
\begin{equation}
\label{crit} A_{nm}=\frac{1}{1+\frac{(n-m)^{2}}{B^{2}}}.
\end{equation}
This model (not yet solved) possesses a fascinating property of
multifractality and is an extremely accurate model for describing
the critical states at the Anderson localization transition point in
dimensionality $d>2$.

\section{Parametrization in terms of
eigenvalues and eigenvectors} Consider a Hermitean $N\times N$matrix
${\bf H}={\bf H}^{\dagger}$. The physical meaning is mostly
contained in the eigenvalues $E_{n}$ of this matrix and also in the
unitary matrix ${\bf U}=({\bf U}^{\dagger})^{-1}$ whose $n$-th
column  is an $n$-th normalized eigenvector ${\bf
\Psi}_{n}\equiv\{\Psi_{n}(r)\}$. Therefore it is sensible to
parametrize  the matrix ${\bf H}$ in the following way:
\begin{equation}
\label{param} {\bf H}={\bf U}\,{\bf E}\,{\bf U}^{\dagger},
\end{equation}
where ${\bf E}={\rm diag}\{E_{n} \}$. Then instead of $N(N+1)/2$
independent entries of the Hermitean matrix ${\bf H}$ one will deal
with $N(N-1)/2$ independent variables of the unitary matrix ${\bf
U}$ plus $N$ eigenvalues.

For invariant ensembles the PDF is independent of the eigenvector
degrees of freedom and is determined only by the eigenvalues. In
particular for the classic WD ensemble it reduces to:
\begin{equation}
\label{meas} P({\bf H})\propto {\rm
exp}\left[-a\sum_{n}E_{n}^{2}\right].
\end{equation}
However the change of variables involves also computing the Jacobian
of the transformation Eq.(\ref{param}). The easiest way of computing
it is to compute the form:
\begin{equation}
\label{form} {\rm Tr}(d{\bf H})^{2}={\bf Tr}\left(2d{\bf f}\,{\bf
E}\,d{\bf f}\,{\bf E}-2{\bf E}^{2}\,(d{\bf f})^{2}+(d{\bf E})^{2}
\right)=2\sum_{n>m}(E_{n}-E_{m})^{2}\,|df_{nm}|^{2}+\sum_{n}(dE_{n})^{2},
\end{equation}
where
$$ d{\bf f}={\bf U}^{\dagger}\,d{\bf U}=-d{\bf U}^{\dagger}\,{\bf U}=-d{\bf
f}^{\dagger}.$$ The set of $f_{nm}$, ($n>m$), are $N(N-1)/2$ natural
"coordinates" related to eigenvectors. Then the Jacobian of the
transformation $d{\bf H}\rightarrow d{\bf f}\,d{\bf E}$ is given by
\begin{equation}
\label{J} J\propto \left\{\begin{matrix}\sqrt{D}, & {\bf
H}&is&real\cr D,& {\bf H}& is &
complex\end{matrix}\right.\;\;\;\;\;\;\;\;\;\;\;D=\Delta^{2}=\prod_{n>m}(E_{n}-E_{m})^{2}.
\end{equation}
\section{Joint probability distribution}
According to Eqs.(\ref{meas}),(\ref{J}) the entire joint probability
distribution function of eigenvalues and eigenvectors for an
arbitrary invariant ensemble Eq.(\ref{inv}) takes the form:
\begin{equation}
\label{JPDF} d{\bf H}P({\bf H})=d{\bf f}d\{E_{n}\}\;{\rm exp}\left[
-\sum_{n=1}^{N} V(E_{n})\right]\, |\Delta|^{\beta},
\end{equation}
where $\Delta$ is the Vandermond determinant:
\begin{equation}
\label{vandermond}  \Delta_{N}=\left|
\begin{matrix}1&1&...&1\cr E_{1}&E_{2}&...&E_{N}\cr
E_{1}^{2}&E_{2}^{2}&...&E_{N}^{2}\cr .& .& ...&.\cr
E_{1}^{N}&E_{2}^{N}&...&E_{N}^{N}\cr
 \end{matrix}\right|=\prod_{n>m} (E_{n}-E_{m}).
\end{equation}
A remarkable property of this distribution is that it is independent
of the eigenvectors and depends only on eigenvalues.

Since the Vandermond determinant is vanishing if any two eigenvalues
coincide $E_{n}=E_{m}$ (two columns of the determinant are equal)
the coincidence of two eigenvalues is statistically improbable. This
is the basic property of the random matrix theory which is called
{\it level repulsion}.

\section{Level repulsion: poor man derivation}
In order to understand the physical origin of level repulsion let us
consider a situation where occasionally  two levels are very close
to each other $|E_{1}-E_{2}|\ll \Delta$, where $\Delta$ is the mean
level separation. Then it is enough to consider only one block of
the random matrix:
$$ \left( \begin{matrix} \varepsilon_{1}& V\cr V^{*}& \varepsilon_{2}\cr\end{matrix}\right)$$
The true energy levels of this two-level system are well known:
\begin{equation}
\label{true}
E_{1,2}=\frac{\varepsilon_{1}+\varepsilon_{2}}{2}\pm\frac{1}{2}\,\sqrt{(\varepsilon_{1}-\varepsilon_{2})^{2}+
|V|^{2}}.
\end{equation}

The two-level correlation function which is the probability density
to find a level at a distance $\omega$ from the given one, is given
by:
\begin{equation}
\label{TLCF} R(\omega)=\int d\varepsilon_{1}d\varepsilon_{2}{\cal
D}V\, \left[
\delta(\omega-\sqrt{(\varepsilon_{1}-\varepsilon_{2})^{2}+
|V|^{2}})-\delta(\omega-\varepsilon_{1}+\varepsilon_{2})\right]\,P(\varepsilon_{1},\varepsilon_{2},
V),
\end{equation}
where
\begin{equation}
\label{DV} {\cal D}V=\left\{\begin{matrix}dV,&\beta=1\cr d\Im V d\Re
V, & \beta=2 \cr
\end{matrix} \right.
\end{equation}
Small energy difference $|E_{1}-E_{2}|\ll \Delta$ implies that {\it
both} $\varepsilon_{1}-\varepsilon_{2}$ and $|V|$ are small. Then
the PDF  $P(\varepsilon_{1},\varepsilon_{2}, V)$ can be considered
independent of $\varepsilon_{1}-\varepsilon_{2}$ and $|V|$. Thus
integrating the $\delta$-functions over
$\varepsilon_{1}-\varepsilon_{2}$ we arrive at:
\begin{equation}
\label{DV1} R(\omega)=\int {\cal D}V\,\left[
\sqrt{1-\frac{|V|^{2}}{\omega^{2}}}-1\right]\,\theta(\omega-|V|).
\end{equation}
Apparently this integral is convergent and the power counting
immediately leads to:
\begin{equation}
\label{pc} R(\omega)\propto\left\{\begin{matrix}\omega, & \beta=1\cr
\omega^{2},&\beta=2\cr
\end{matrix} \right.
\end{equation}

The simple analysis above illustrates two important points. One
--physical-- is that the level repulsion is nothing but the {\it
avoided level crossing} which is well known in quantum mechanics and
which is caused by the $|V|^{2}$ term in the square root in
Eq.(\ref{true}). The other one -- formal-- is that the pseudo-gap in
$R(\omega)$ near $\omega=0$ is the effect of the phase volume ${\cal
D}V$ and the power of $\omega$ depends on the number of independent
components of $V$ which is 1 in case $V$ is a real number and 2 if
$V=\Re V + i \Im V$ is a complex number.

It is known that the algebra of real and complex numbers allows only
one further step of generalization. This is the algebra of
quaternions:
\begin{equation}
\label{quat} \tau_{0}=1,\;\;\;\;\tau_{1,2,3}=i\sigma_{1,2,3},
\end{equation}
where $\sigma_{1,2,3}$ are $2\times 2$ Pauli matrices. It appears
that this generalization makes sense in the context of random
matrices too. Namely, one can consider random matrices which entries
are  {\it real quaternions}, i.e.:
\begin{equation}
\label{V-quat} V=\sum_{i=0}^{3}\xi_{i}\,\tau_{i},
\end{equation}
with real components $\xi_{i}$. This generalization corresponds to
$\beta=4$ in Eqs.(\ref{Psi-q},\ref{CSM}).
\section{Time-reversal symmetry and the Dyson symmetry classes}
It turns out the the parameter $\beta$ is related with the
time-reversal symmetry. In order to see this we note that the
time-reversal operator ${\cal T}$ should obey a basic property
$${\cal T}^{2}=\alpha\,1,\;\;\;\;\;\;\;|\alpha|=1$$ (time reversal applied twice leaves the
wave function unchanged). As the time reversal operator should
involve the complex conjugation of wave function one may write:
\begin{equation}
\label{cc} {\cal T}=K\,C,
\end{equation}
where $C$ is the complex conjugation operator and $K$ is an operator
such that
\begin{equation}
\label{T2} K\,C\,K\,C=K\,K^{*}=\alpha\,1.
\end{equation}
But $K$ must be a unitary operator (as the norm of the wave function
must be conserved). That is why
\begin{equation}
\label{uni} K^{*}\,K^{T}=1.
\end{equation}
From these two conditions one finds:
\begin{equation}
\label{al}
K=\alpha\,K^{T}=\alpha\,(\alpha\,K^{T})^{T}=\alpha^{2}\,K.
\end{equation}
Thus we conclude that
\begin{equation}
\label{alph} \alpha^{2}=1\;\;\;\;\;\;\Rightarrow \alpha=\pm1.
\end{equation}

 For spinless particles (or particles with even spin) we have:
\begin{equation}
\label{even} {\cal T}^{2}=1,
\end{equation}
and $K$ can always be chosen to be a unity operator $K=1$.

  However, for particles with half-integer spin
\begin{equation}
\label{odd} {\cal T}^{2}=-1,
\end{equation}
and $K$ is not an identity operator. In particular for
spin-$\frac{1}{2}$ particles $K$ is a $2\times 2$ matrix. Using
Eqs.(\ref{T2}),(\ref{uni}) one can show that up to a phase factor
$e^{i\theta}$ the matrix $K$ is equal to:
\begin{equation}
\label{K} K=\left(\begin{matrix}0 & 1\cr -1 & 0 \cr  \end{matrix}
\right)
\end{equation}
The physical meaning of this operator is very simple: it flips the
spinor.

The time-reversal symmetry $${\cal T}H{\cal T}^{\dagger}=H$$ in the
cases Eq.(\ref{even}) and Eq.(\ref{odd}) implies, respectively
\begin{equation}
\label{even1} H=H^{*}
\end{equation}
and
\begin{equation}
\label{even1} H=-K\,H^{*}\,K=K\,H^{*}\,K^{\dagger}.
\end{equation}
In the first case time reversal symmetry requires the Hamiltonian
matrix ${\bf H}$ to be {\it real}, which corresponds to $\beta=1$.
In the second case one can do a simple algebra exercise and show
that the condition Eq.(\ref{even1}) is fulfilled if the Hamiltonian
matrix ${\bf H}$ has entries of the form Eq.(\ref{V-quat}) with {\it
real} coefficients $\xi_{i}$. As was already mentioned this case
corresponds to $\beta=4$.

It is remarkable that Eq.(\ref{even1}) leads to a two-fold
degeneracy of energy levels known as the Kramers degeneracy. To
prove this statement we assume that the wave vector $\psi$
corresponds to the eigenstate with the energy E, i.e
$$
H\,\psi=E\,\psi,\;\;\;\;H^{*}\,\psi^{*}=E\,\psi^{*}.
$$
Multiplying the second of these equations by $K$ and using
Eq.(\ref{even1})one obtains:
\begin{equation}
K\,H^{*}\,\psi^{*} = -
K\,H^{*}\,K\,(K\,\psi^{*})=+H\,(K\,\psi^{*})=E\,(K\,\psi^{*}).
\end{equation}
The last equality implies that $K\,\psi^{*}$ is also an eigenvector
corresponding the eigenvalue $E$. The two eigenvectors are
different. Indeed, if $K\,\psi^{*}=\lambda\,\psi$ then
$K^{2}\,\psi=\lambda^{*}\,K\,\psi^{*}=|\lambda^{2}|\,\psi=-\psi$ and
$\psi$ must be zero. This completes the proof of the two-fold
degeneracy.

Thus we see that the case $\beta=1$ (known as the {\it orthogonal
ensemble}) corresponds to the particles with an even spin and a
Hamiltonian that preserves the time reversal symmetry. The case
$\beta=4$ (known as the {\it symplectic ensemble})corresponds to
particles with an odd spin and a {\it spin-dependent} Hamiltonian
that preserves time reversal symmetry. In particular it applies to a
system with the {\it spin-orbit} interaction. The case $\beta=2$
(known as the {\it unitary ensemble})does not assume any definite
relationship between $H$ and $H^{*}$, and thus the time-reversal
symmetry must be broken (e.g. by magnetic field or magnetic
impurities).

According to the initial idea of Wigner and Dyson all systems with
complex interactions do not possess any symmetry but possibly
time-reversal symmetry and thus should be classified according to
one of the three symmetry classes discussed above.
\section{Extension of the Dyson symmetry classes}
It has been discovered relatively recently (Altland and Zirnbauer,
1995) that the Dyson list of symmetry classes can be naturally
extended from 3 to 10 symmetry classes if one  introduces, in
addition to the time-reversal symmetry, also the {\it particle-hole}
symmetry. This requires a certain quasi-relativistic description
where there exist both particles and anti-particles (holes). In
application to condensed matter physics such situation realizes in
superconductivity which basic description (the Bogolyubov-de Gennes
equation) is in terms of the two coupled Schroedinger-type equations
for particles and holes. The block-matrix form of such a Hamiltonian
reads as follows:
\begin{equation}
\label{BdG} H=\left( \begin{matrix} {\cal H}& \Delta\cr
\Delta^{\dagger}&-{\cal H}^{*}
\end{matrix}\right).
\end{equation}
The off-diagonal part $\Delta$ could be even or odd under the
transposition:
\begin{equation}\label{delta-eo}
\Delta^{T}=\pm \Delta.
\end{equation}
The first choice (sign $+$) corresponds to the {\it singlet}
superconductive paring of fermions, which is odd under spin
permutations and thus must be even under orbital permutation. The
second choice (sign $-$) corresponds to the triplet
superconductivity which is odd under orbital permutation.

 This Hamiltonian acts on a wave function
\begin{equation}
\label{wave} \Psi=\left(\begin{matrix}p\cr h
\end{matrix} \right),\;\;\;p=\left( \begin{matrix} p\uparrow \cr p\downarrow\end{matrix}\right),\;\;
h=\left( \begin{matrix} h\uparrow \cr
h\downarrow\end{matrix}\right).
\end{equation}
Note that here the block $2\times 2$ matrix in the {\it
particle-hole} space which should not be confused with the spinor
space we considered in connection with the time reversal symmetry in
the previous section. Thus inclusion of {\it both} time-reversal and
the particle-hole symmetry requires to consider the product Hilbert
space of spinor and particle-hole components of the wave functions.
One can check that the Hamiltonian of the form Eq.(\ref{BdG}) obeys
the symmetry relation:
\begin{equation}\label{phs-s} H=-L\,H^{*}\,L^{\dagger}, \end{equation}
where for the singlet paring \begin{equation} \label{P}L_{2}=\left(
\begin{matrix} 0 & 1\cr -1&0\end{matrix}\right)_{ph},\;\;\;L_{2}^{2}=-1.
\end{equation}
and for the triplet paring:
 \begin{equation} \label{P}L_{1}=\left( \begin{matrix} 0 & 1\cr
1&0\end{matrix}\right)_{ph},\;\;\;L_{1}^{2}=+1
\end{equation}

Eq.(\ref{phs-s}) is the same type of constraint as Eq.(\ref{even1})
but with the opposite sign of the r.h.s. This change of sign
reflects the fact that the charge-conjugation operator ${\cal C}$ is
{\it anti-unitary}. The two cases of the singlet and triplet paring
with the corresponding behavior
\begin{equation}
\label{K-2} L^{2}=\alpha,\;\;\;\;\alpha=\pm 1
\end{equation}
are analogous to the two realizations of the time-reversal
transformations discussed in the previous section. However, the
consequences of the time-reversal symmetry (TRS) and the
particle-hole (PH) symmetry for the spectrum of random matrices are
different: the TRS with ${\cal T}^{2}=-1$ implies the Kramers
degeneracy (each level is doubly degenerate) while PH symmetry with
$K^{2}=-1$ implies that the spectrum is exactly symmetric with
respect to $E=0$ (for each level $E_{n}>0$ there is a corresponding
level $E_{m}=-E_{n}<0$). This is related with the change of sign in
Eq.(\ref{phs-s}) compared to Eq.(\ref{even1}).

One may ask a question about the states at $E=0$. Clearly, if the
matrix size is even as in Eq.(\ref{BdG}), the total number of states
is even too. Every non-zero eigenvalue of the Hamiltonian
Eq.(\ref{BdG}) enters in pairs $(E,\;-E)$ which corresponds to a
pair of {\it different} states $(\psi, \; L\psi^{*}$. At $E=0$ it is
not guaranteed that the states $\psi$ and $L\psi^{*}$ are really
different. In order to clarify this issue one should repeat the
algebra presented at the end of the previous section but for the
case of the particle-hole symmetry. Indeed,
\begin{equation}
\label{assumpt} L\,\psi^{*}=\lambda\,\psi, \;\;\;\;\;|\lambda|=1
\end{equation}
would mean
\begin{equation}
\label{consenq} L L^{*}\psi
=L^{2}\,\psi=\lambda^{*}\,L\psi^{*}=|\lambda|^{2}\,\psi=\psi.
\end{equation}
For the singlet case $L_{2}^{2}=-1$, and Eq.(\ref{consenq}) cannot
be satisfied. This means that the two states $\psi$ and ${\cal
C}\psi\equiv L_{2}\psi^{*}$ are indeed different. However, for the
triplet paring $L_{1}^{2}=+1$ and the assumption ${\cal C}\psi=\psi$
does not lead to a contradiction. Clearly, this may only happen at
zero energy, as otherwise the same states would lead to different
energies $E$ and $-E$. Thus in the case of triplet paring one may
have a state at zero energy that is equal to its particle-hole
conjugated. There should be obviously even number of such states,
e.g. $\psi_{1}={\cal C}\psi_{1}$ and $\psi_{2}={\cal C}\psi_{2}$, as
the total number of states is even. As the particle-hole
transformation ${\cal C}$ transforms $c^{\dagger}$ into $c$ and vise
versa, the state $\psi={\cal C}\psi$ corresponds to the {\it
Majorana fermion}.

Now return to the symmetry classes extension. If we denote the
behavior of the system with respect to each of the two the symmetry
transformations as $0$ (no symmetry), $+1$ ($\alpha=+1$) and $-1$
($\alpha=-1$) then we obtain $3\times 3=9$ possible combinations of
$(p,p')$, (where $p,p'=0,\pm 1$) and 9 respective symmetry classes.
The 10-th class appear because when neither of the two symmetries is
present (the $(0,0)$ case) the symmetry with respect to their
product ${\cal T}{\cal C}$ may be present or absent. So the $(0,0)$
case (and mathematic says that only this case) is actually split
into two classes.

The symmetry under the product ${\cal T}{\cal C}$ (with ${\cal
T}^{2}=-1$) can always be cast (in some special basis) as a symmetry
constraint:
\begin{equation}
\label{chir} H=-\Sigma_{z}\,H\,\Sigma_{z}.
\end{equation}

Indeed, Combining Eqs.(\ref{phs-s}) and (\ref{even1}) one obtains:
\begin{equation}
\label{expl} H=-M\,H\,M,
\end{equation}
where the $4\times 4$ matrix $M$ takes one of two forms:
$$
\left(\begin{matrix} 0&0&0&1\cr 0&0&1&0\cr 0& -1 &0&0\cr -1&0&0&0\cr
\end{matrix}\right),\;\;\;\;\;\;\left(\begin{matrix} 0&0&0&1\cr 0&0&-1&0\cr 0& -1 &0&0\cr 1&0&0&0\cr
\end{matrix}\right).
$$
The two $+1$ and two $-1$ can be arranged by permutation of columns
in the standard way:
$$
\Sigma_{z}=\left(\begin{matrix} 1&0&0&0\cr 0&1&0&0\cr 0& 0 &-1&0\cr
0&0&0&-1\cr
\end{matrix}\right)
$$
The corresponding $H$ obeying Eq.(\ref{expl}) is block-off-diagonal:
$$
H=\left(\begin{matrix} 0&0&h_{1}&h_{2}\cr 0&0&h_{3}&h_{4}\cr h_{1}&
h_{3} &0&0\cr h_{2}&h_{4}&0&0\cr
\end{matrix}\right)
$$

 This is a new type of the symmetry, which symmetry constraint
$Eq.(\ref{chir})$ contains $H$ (rather than $H^{*}=H^{T}$) in the
r.h.s.. It is the simple consequence of the product ${\cal T}{\cal
C}$, where the symmetry constrains Eq.(\ref{even1}),(\ref{phs-s})
for ${\cal T}$, ${\cal C}$
  contain $H^{*}$ in the corresponding
r.h.s.

The symmetry under ${\cal L}={\cal T}{\cal C}$ is known as the {\it
sublattice}, or {\it chiral} symmetry. The reason for the first
nickname is that the off-diagonal structure of Eq.(\ref{block-off})
appears in one of the simplest models of disorder: the
one-dimensional chain with on-site energies $\varepsilon_{n}=0$ and
the hopping to the nearest neighbor $t_{n,n\pm 1}$ containing a
random part. If one introduces two sublattices $A$ (containing even
sites ) and $B$ (containing odd sites), then the random hopping will
connect only different sublattices resulting in the block
off-diagonal terms $H_{AB}=V_{AB}$. There will be no terms $H_{AA}$
or $H_{BB}$. Indeed, the diagonal entries of $H_{AA}$ and $H_{BB}$
are zero because the on-site energy is zero while the off-diagonal
entries are zero due to the absence of the hopping integrals other
than between the nearest neighbors (which belong to different
sublattices).

The presence of the chiral symmetry usually favors delocalization.
For instance in the one-dimensional disordered chain discussed above
the localization radius tends to infinity as the energy of the
eigenstate approaches zero.

\section{Level repulsion: classical and quantum analogy}
\subsection{Classical plasma with logarithmic interaction}

 One that for the Gaussian invariant ensemble one can rewrite the
JPDF Eq.(\ref{JPDF}) in the following way:
\begin{equation}
\label{Hamil} J\,P({\bf H})\propto {\rm exp}[-\beta\,{\cal
L}],\;\;\;\;\;\;\;\;\;{\cal
L}=-\sum_{n>m}\ln|E_{n}-E_{m}|+a'\sum_{n}E_{n}^{2},
\end{equation}
where we introduces the Dyson symmetry parameter:
\begin{equation}
\label{beta} \beta=\left\{\begin{matrix}1, & for & real &symmetric&
{\bf H}\cr 2, & for & complex & Hermitean & {\bf H}\cr 4, &for
&real-quaternionic & Hermitean &{\bf H}\cr
\end{matrix} \right.
\end{equation}
Note that by a proper choice of energy units the parameter
$a'=a\beta^{-1}$ can be set equal to $\frac{1}{2}$ which will be
always assumed throughout the lecture notes. Thus there is only one
important parameter $\beta$ in the classic WD random matrix theory.

Looking at Eq.(\ref{Hamil}) one concludes that the PDF in the $({\bf
f},\,{\bf E})$ representation coincides with the partition function
of classical particles repelling each other {\it logarithmically},
in a harmonic {\it confinement potential}. The Dyson symmetry
parameter $\beta$ plays a role if an inverse temperature.

The above derivation which lead to Eq.(\ref{Hamil}) can be repeated
for an arbitrary invariant RME. The corresponding energy functional
${\cal L}$ of the logarithmically repelling particles will differ
from Eq.(\ref{Hamil}) only by the confinement potential which will
be no longer harmonic but rather $\beta^{-1}\,V(E_{n})$. The basic
property of the PDF which dependends only on the set of eigenvalues
$E_{n}$ (but not the eigenvector variables ${\bf f}$) is retained
for all the invariant RME making the corresponding eigenfunction
statistics trivial.

This is no longer true once the invariance under basis rotation is
broken. The latter circumstance is what makes non-invariant
ensembles difficult to solve but at the same time having a rich
variety of eigenfunction statistics.

\subsection{Quantum analogy}
Besides the analogy with logarithmically repelling {\it classical}
particles at {\it finite temperature} $\beta^{-1}$ living in one
dimension (1d) there is also an important analogy with the system of
{\it quantum} particles in 1d. To facilitate this analogy let us
remind that the Jacobian in Eq.(\ref{J}) can be expressed as the
power of the Vandermond determinant Eq.(\ref{vandermond})
\begin{equation}
\label{J-vandermond} J\propto |\Delta|^{\beta}.
\end{equation}
The property of the Vandermond determinant is that
\begin{equation}
\label{VdM-der} \sum_{n=1}^{N}\frac{\partial^{2}\Delta_{N}}{\partial
E_{n}^{2}}=0.
\end{equation}
Another property is that it changes sign upon any permutation of two
$E_{n}$ and $E_{m}$.

These two properties imply that $\Delta_{N}$ can be considered as
the many-body wave function $\Psi(\{ E_{n}\})$ of the system of $N$
free fermions with the Hamiltonian consisting only of kinetic
energy:
\begin{equation}
\label{ham-ferm} {\cal
H}_{K}=-\sum_{n=1}^{N}\frac{\partial^{2}}{\partial E_{n}^{2}}
\end{equation}
Moreover, as the energy of the corresponding many-body state is
minimal possible for kinetic energy ${\cal E}=0$, this is a {\it
ground state} of this free fermionic system.

So we come to the statement that the Jacobian Eq.(\ref{J}) at
$\beta=2$ is the  probability density for the ground state of the
free fermion system in the entire space.
\begin{equation}
\label{Psi-q} J\sim
|\Psi_{0}(E_{n})|^{2}\propto\prod_{n>m}|E_{n}-E_{m}|^{\beta}.
\end{equation}
Note that the system of fermions in an infinite space is not well
defined, as it expands indefinitely. Formally this is seen from the
fact that the wave function Eq.(\ref{Psi-q}) is not normalizable. In
order to fix this pathology one has to consider a full probability
distribution function Eq.(\ref{JPDF}) which includes also the
confinement potential $V(E_{n})$.

For the harmonic confinement potential the property
Eq.(\ref{VdM-der}) can be generalized in the following way:
\begin{equation}
\label{VdM-der-gen}
\frac{1}{2m}\sum_{n=1}^{N}\frac{\partial^{2}}{\partial
E_{n}^{2}}\left[\Delta_{N}\,e^{-\frac{m}{2}\sum_{n=1}^{N}E_{n}^{2}}\right]
=\left[-{\cal
E}_{N}+\frac{m}{2}\sum_{n=1}^{N}E_{n}^{2}\right]\,\Delta_{N}\,
e^{-\frac{m}{2}\sum_{n=1}^{N}E_{n}^{2}}.
\end{equation}
Now we see that
\begin{equation}
\label{gen-wf} \Psi_{0}\propto
\Delta_{N}\,e^{-\frac{m}{2}\sum_{n=1}^{N}E_{n}^{2}},
\end{equation}
is an eigenfunction of the free fermions with mass $m$ in a harmonic
confinement potential $V(E)=\frac{m}{2}E^{2}$. It corresponds to a
certain positive energy ${\cal E}_{N}$ which arises due to
confinement of fermions.

Now suppose that this property is valid also for arbitrary $\beta$
and check that the wave function
\begin{equation}
\label{wwf}
\Psi_{0,\beta}(\{E_{n}\})=\prod_{n>m}|E_{n}-E_{m}|^{\beta/2}\,{\rm
sgn}(E_{n}-E_{m})\,e^{-\frac{m}{2}\sum_{n=1}^{N}E_{n}^{2}}
\end{equation}
is the eigenfunction of a certain Hamiltonian. Note that the
coefficient $m$ can be done arbitrary small by a proper choice of
$E_{n}$ units. So, for simplicity of further derivation we consider
the case $m\rightarrow 0$.

To this end we take the sum of second derivatives of the wave
function applying the kinetic energy operator Eq.(\ref{ham-ferm}) to
Eq.(\ref{wwf}) with $m\rightarrow 0$. The result appears to be
proportional to $\Psi_{0,\beta}(\{E_{n}\})$:
\begin{equation}
\label{pot} -{\cal H}_{K}\,\Psi_{0,\beta}=\frac{\beta}{2}\left(
\frac{\beta}{2}-1\right)\,|\Delta_{N}|^{-2}\sum_{n=1}^{N}\left(
\frac{\partial\Delta_{N}}{\partial
E_{n}}\right)^{2}\,\Psi_{0,\beta}.
\end{equation}
Thus at any $\beta\neq 2$ the system of fermions equivalent to an
invariant random matrix theory is interacting with the interaction
Hamiltonian:
\begin{equation}
\label{int-Ham} {\cal H}_{{\rm int}}=\frac{\beta}{2}\left(
\frac{\beta}{2}-1\right)\,|\Delta_{N}|^{-2}\sum_{n=1}^{N}\left(
\frac{\partial\Delta_{N}}{\partial E_{n}}\right)^{2}.
\end{equation}
Now if we use the property of the Vandermond determinant:
\begin{equation}
\sum_{n=1}^{N}\left(\frac{\partial \Delta_{N}}{\partial
E_{n}}\right)^{2}=2|\Delta_{N}|^{2}\,\sum_{n=1}^{N}\frac{1}{(E_{n}-E_{m})^{2}}
\end{equation}
we finally obtain the total Hamiltonian of an equivalent system of
fermions:
\begin{equation}
\label{CSM}
\hat{H}=-\frac{1}{2}\sum_{n=1}^{N}\frac{\partial^{2}}{\partial
E_{n}^{2}}+\frac{\beta}{2}\left(\frac{\beta}{2}-1
\right)\,\sum_{n>m}^{N}\frac{1}{(E_{n}-E_{m})^{2}}.
\end{equation}
This is the celebrated Calogero-Sutherland Hamiltonian
\cite{CalSuth} with the inverse square interaction. For $\beta=2$
the interaction constant vanishes and the entire level repulsion is
due to fermionic nature of the fictitious particles. For $\beta=1$
there is some attraction on top of the free fermionic mutual
avoiding, while for $\beta=4$ the interaction is repelling. This
additional interaction explains why the level repulsion for
$\beta=4$ is stronger then for $\beta=2$ and for $\beta=1$ it is
weaker than for $\beta=2$.

\section{Plasma model and the Wigner semi-circle}
The model of classical particles in one dimension with logarithmic
repulsion Eq.(\ref{Hamil}) can be represented by a continuous energy
functional:
\begin{equation}
\label{densityFun} {\cal L}=-\frac{1}{2}\,\int dE\int
dE'\,\varrho(E)\,\varrho(E')\,\ln|E-E'|+\frac{1}{\beta}\int
dE\;\varrho(E)\,V(E),
\end{equation}
expressed through the exact density
\begin{equation}
\label{ex-dens} \varrho(E)=\sum_{n}\delta(E-E_{n}).
\end{equation}
Now we make two {\it assumptions}:\\
(i) replace $\varrho(E)$ by an {\it ensemble average} value
$\rho(E)$ and\\
(ii) neglect the thermal fluctuations by minimizing the energy
functional Eq.(\ref{densityFun}) (with $\varrho$ replaced by $\rho$)
 instead of computing the partition function
\begin{equation}
\label{part} \sum_{{\rm config.}\{E_{n}\}}{\rm e}^{-\beta\,{\cal
L}}.
\end{equation}
As a result one gets a kind of {\it mean field} approximation which
is justified by the long-range, logarithmic nature of interaction.

Minimizing Eq.(\ref{densityFun}) with respect to $\rho(E)$ and
differentiating both sides with respect to $E$ one obtains:
\begin{equation}
\label{plasEq} \int_{-\infty}^{+\infty}
\rho(E')\,\frac{dE'}{E-E'}=\frac{1}{\beta}\frac{dV}{dE}\equiv f(E).
\end{equation}
The physical meaning of this equation is very simple: the force
acting upon the given "particle" from all other particles should be
balanced by the confining force. This is the condition of the plasma
equilibrium.

From the mathematical viewpoint Eq.(\ref{plasEq}) is a {\it strongly
singular} integral equation. Its solution is well known
\cite{Muskh}. For an even function $V(E)=V(-E)$ it reads:
\begin{equation}
\label{Muskh}
\rho_{0}(E)=\frac{1}{\pi^{2}}\,\sqrt{D^{2}-E^{2}}\int_{-D}^{D}\frac{f(E')}{\sqrt{D^{2}-E'^{2}}}
\,\frac{dE'}{E'-E},
\end{equation}
where the principle value of the integral is assumed in
Eq.(\ref{plasEq}) and Eq.(\ref{Muskh}), namely
\begin{equation}
\label{valp} \frac{1}{E'-E}\rightarrow
\frac{1}{2}\,\left(\frac{1}{E'-E-i0}+\frac{1}{E'-E+i0} \right).
\end{equation}
This definition allows to make an analytic continuation of
Eq.(\ref{Muskh}) for $E$ in the complex plane with the cut along the
real axis  with $|E|>D$, where the bandwidth $D$ should be chosen
from the condition that the total number of eigenvalues is equal to
the size of matrix $N$:
\begin{equation}
\label{norm-cond} \int_{-D}^{D}\rho_{0}(E)\,dE = N.
\end{equation}
Namely, $\rho(E)$ can be represented as a sum of  a function
$\rho_{+}(E)$ which is regular in the upper half-plane $\Im E>0$ and
a function $\rho_{-}(E)$ which is regular in the lower half-plane
$\Im E<0$:
\begin{equation}
\label{anal-cont}
\rho_{0}(E)=\rho_{+}(E)+\rho_{-}(E),\;\;\;\;\;\rho_{\pm}(E)=
\frac{1}{2\pi^{2}}\,\sqrt{D^{2}-E^{2}}\int_{-D}^{D}\frac{f(E')}{\sqrt{D^{2}-E'^{2}}}
\,\frac{dE'}{E'-E\mp i 0}.
\end{equation}
It is important that  along the cut $|E|>D$ the analytic function
$\sqrt{D^{2}-E^{2}}=\pm i\sqrt{E^{2}-D^{2}}$ has different signs
just above and just below the cut. This means that for $|E|>D$
\begin{equation}
\label{pm}
\rho_{+}(E)+\rho_{-}(E)=-\frac{1}{\pi}\,\sqrt{E^{2}-D^{2}}\int_{-D}^{D}dE'\,\frac{f(E')}{\sqrt{D^{2}-E'^{2}}}
\,\delta(E-E')=0.
\end{equation}
On the other hand, for  real $E$ beyond the cut ($|E|<D$) one
obtains:
\begin{equation}
\label{pm1} \rho_{+}(E)-\rho_{-}(E)=\frac{2\pi
i}{2\pi^{2}}\,\sqrt{D^{2}-E^{2}}\int_{-D}^{D}\frac{f(E')}{\sqrt{D^{2}-E'^{2}}}
\,\delta(E-E')=\frac{i}{\pi}\,f(E).
\end{equation}
Now we are in a position to check that Eq.(\ref{Muskh}) is really a
solution of Eq.(\ref{plasEq}) for real $E$ beyond ($|E|<D$) the cut.
Indeed, the integral over the real axis in Eq.(\ref{plasEq}) can be
closed either through the upper complex half-plane of $E'$ or
through the lower half-plane. We use the first option for the part
containing $\rho_{+}(E')$ and the second option for the part
containing $\rho_{-}(E')$. Each of the two contour integrals allows
for the evaluation using the residue theorem. Then omitting the
terms which do not have poles in the corresponding half-plane we
obtain:
$$
\int_{-\infty}^{+\infty}
\rho_{0}(E')\,\frac{dE'}{E-E'}=\frac{1}{2}\int_{{\rm
upper}}\frac{\rho_{+}(E')}{E-E'+i0}\,dE'+\frac{1}{2}\int_{{\rm
lower}}\frac{\rho_{-}(E')}{E-E'-i0}\,dE'=-\pi
i\,[\rho_{+}(E)-\rho_{-}(E)]=f(E).
$$
This concludes the proof that Eq.(\ref{Muskh}) is indeed a solution
of the integral equation Eq.(\ref{plasEq}). The beauty of the proof
is that it is based only on the analytic properties of the solution.

For the Gaussian ensemble where $f(E')=E'$ the integral in
Eq.(\ref{Muskh}) is actually independent of $E$ ({\bf show this
using the definition of the principle value of the integral})
\begin{equation}
\label{int-ind} \int_{-D}^{D}\frac{E'}{\sqrt{D^{2}-E'^{2}}}
\,\frac{dE'}{E'-E}=\int_{-D}^{D}\frac{1}{\sqrt{D^{2}-E'^{2}}}
\,dE'=\pi.
\end{equation}
and the average density is the celebrated {\it semi-circle}:
\begin{equation}
\label{semi-circ} \rho_{0}(E)=\frac{1}{\pi}\,\sqrt{2N-E^{2}}.
\end{equation}
\section{Probability of having a hole in spectrum and the Wigner Surmise}
One of the most popular statistics of eigenvalues of complex quantum
systems is the {\it the level spacing distribution} $P(\omega)$: the
probability density to have a level at a distance $\omega$ from a
given level and no other levels between them. For $\omega$ much
smaller than the mean level spacing $\Delta=\rho^{-1}$, it is
improbable that in between of the two close levels there is yet
another one or several levels. Then the requirement of having no
levels in between of the two is unimportant and the leading term in
$P(\omega)$ is the same as in the two-level correlation function
$R(\omega)\propto \omega^{\beta}$ at $\omega\ll \Delta$. However,
for $\omega\gg \Delta$ the two statistics dramatically differ:
$R(\omega)$ tends to a constant whereas $P(\omega)$ is very small
due to a small probability to have no levels in between of the two
levels separated by a large distance. Basically the $P(\omega)$ for
$\omega\gg \Delta$ is limited by the probability of having a hole of
the size $\omega$ in the spectrum. Let us find this probability
using the plasma analogy.

As for any fluctuation, the probability of having a hole is given by
the energy cost $\delta {\cal L}$ of this configuration relative to
the equilibrium one:
\begin{equation}
\label{fluc} P(\omega)\propto {\rm exp}(-\beta\,\delta{\cal L}).
\end{equation}
One can cast the energy difference in the following way:
\begin{equation}
\label{deltaH} \Delta {\cal
L}=\frac{1}{2}\int_{C}dE\int_{C}dE'\;\delta\rho(E)\,\delta\rho(E')\,\ln|E-E'|-\frac{1}{2}
\int_{-\omega/2}^{\omega/2}dE\int_{-\omega/2}^{\omega/2}dE'\;\rho_{0}(E)\,\rho_{0}(E')\,\ln|E-E'|,
\end{equation}
where the integrals in the first term run over the real axis {\it
outside the gap region} and in the second term they run {\it over
the gap region}; $\rho_{0}(E)$ is the equilibrium density without
the gap and $\delta\rho(E)=\rho_{\omega}(E)-\rho_{0}(E)$ with
$\rho_{\omega}(E)$ being the solution of the integral equation
Eq.(\ref{plasEq}) with the {\it additional condition} that there is
a gap for $|E|<\omega/2$.

The solution with the gap can also be constructed and and looks as
follows:
\begin{equation}
\label{gapSol}
\rho_{\omega}(E)=\frac{2|E|}{\pi^{2}\,\sqrt{E^{2}-(\omega/2)^{2}}}\,\sqrt{\bar{D}^{2}-E^{2}}\,
\int_{\omega/2}^{\bar{D}}
\frac{f(E')}{\sqrt{\bar{D}^{2}-E'^{2}}}\,\frac{\sqrt{E'^{2}-(\omega/2)^{2}}}{E'^{2}-E^{2}}\,dE'.
\end{equation}
It is important that for the {\it steep confinement} $V(E)\sim
|E|^{\alpha}$ ($\alpha>1$) there is a scale separation, namely the
integral in Eq.(\ref{gapSol}) varies slowly as a function of $E$
with the typical scale of $D\sim N^{1/\alpha}$. In the large $N$
limit one can disregard this dependence and consider
\begin{equation}
\label{DoS}
\rho_{\omega}(E)=\rho_{0}\,\frac{|E|}{\sqrt{E^{2}-(\omega/2)^{2}}}.
\end{equation}
One can immediately recognize the gapped density of states with the
square-root divergency near the gap edges similar to the one for a
BCS superconductor.

Now by making a re-scaling $E\rightarrow s x$, $E'\rightarrow s x'$
and observing that the double integral in the first term is
convergent for $\rho_{\omega}(E)$ of the form Eq.(\ref{DoS}) we
immediately obtain that $\Delta{\cal L}=\omega^{2}\,(a+b\ln
\omega)$. More detailed inspection show that the coefficient $b=0$.

Indeed, the coefficient $b$ is proportional to
\begin{equation}
\label{insp} \left(\int_{{\cal
C}}\delta\rho(E)\,dE\right)^{2}-\left( \int_{-\omega/2}^{\omega/2}
\rho_{0}\,dE\right)^{2}.
\end{equation}
On the other hand, the conservation of the total number of levels
requires:
\begin{equation}
\label{cons} \int_{{\cal C}}\delta\rho(E)\,dE =
\int_{-\omega/2}^{\omega/2}\rho_{0}.
\end{equation}
Raising the l.h.s. and the r.h.s. of the last equation to second
power one proves the statement $b=0$.

 The final result for
$\Delta{\cal L}$ reads:
\begin{equation}
\label{delta-L} \Delta{\cal
L}=\frac{\pi^{2}}{16}\,(\rho_{0}\,\omega)^{2}\approx
0.62\,(\rho_{0}\,\omega)^{2}.
\end{equation}
This implies that the spacing distribution function for large level
separations $s=(\omega/\Delta)\gg 1$ is given by:
\begin{equation}
\label{asym-P} P(s)\propto {\rm exp}\left(-\frac{\pi^{2}
\beta}{16}\,s^{2}\right).
\end{equation}
Note that the "Gaussian" form of $P(s)$ has nothing to do with the
quadratic confinement potential (Gaussian invariant ensemble). In
fact $P(s)$ has the same asymptotic form  Eq.(\ref{asym-P})for all
{\it steep} confinement potentials.\\
 {\bf check that it has the
same form for the confinement potential $V(E)=E^{4}$}.
\\
Finally we mention a famous {\it interpolation formula} for $P(s)$
known as the $Wigner Surmise$:
\begin{equation}
\label{WS} P(s)=A(\beta)\,s^{\beta}\,{\rm
exp}\left[-B(\beta)\,s^{2}\right],\;\;\;\;\;s=\frac{\omega}{\Delta}.
\end{equation}
The coefficients $A(\beta)$ and $B(\beta)$ are found from two
conditions: the normalization to the total probability 1 and the
condition that the mean level spacing in the units of $s$ is one:
\begin{equation}
\label{WigSur}
\int_{0}^{\infty}P(s)\,ds=1,\;\;\;\;\;\;\int_{0}^{\infty}s\;P(s)\,ds=1.
\end{equation}
These conditions result in:
\begin{equation}
\label{A}
A(\beta)=2B(\beta)^{\frac{\beta}{2}+1}\,\Gamma\left(\frac{\beta}{2}+1
\right),\;\;\;\;\;B(\beta)=\left\{\begin{matrix}\frac{\pi}{4}\approx
0.78\;(0.62)&\beta=1\cr \cr\frac{4}{\pi}\approx 1.27\;(1.24) &
\beta=2\cr \cr\frac{64}{9\pi}\approx 2.26\,(2.48) & \beta=4 \cr
\end{matrix} \right.
\end{equation}
For comparison we give in the brackets the exact values of $B_{{\rm
exact}}(\beta)=\pi^{2}\beta/16$. One can see that they are rather
close to the approximate values of the Wingner Surmise, especially
for $\beta=2$.

 \section{Level compressibility, normalization sum rule and normalization anomaly}
The {\it two-level correlation function} (TLCF) is formally defined
as a correlation function of the exact density of states
Eq.(\ref{ex-dens}):
\begin{equation}
\label{forTLCF} R_{N}(E,E')=\frac{\langle\varrho(E)\,\varrho(E')
\rangle}{\rho(E)\,\rho(E')}\equiv
\rho^{-1}(E)\delta(E-E')+1+Y_{N}(E,E').
\end{equation}
The $\delta$-function in Eq.(\ref{forTLCF}) is the self-correlation
coming from one and the same level $n$ in the sum in
Eq.(\ref{ex-dens}). The 1 term gives the asymptotic value of TLCF at
energy separations $|E-E'|\gg \Delta$ when the average of two
densities of states can be decoupled. The function $Y(E,E')$ gives
then a regular contribution to the TLCF which decreases to zero as
$|E-E'|$ increase.

There is an important {\it normalization sum rule} that applies to
the TLCF. Indeed, consider
\begin{equation}
\label{step1} \int dE'\,\rho(E)\,\rho(E')\,R_{N}(E,E')=\int
dE'\,\langle\varrho(E)\,\varrho(E') \rangle=\langle\varrho(E)\,\int
dE'\,\varrho(E') \rangle
\end{equation}
The total number of states is equal to the number of degrees of
freedom $N$ and {\it does not fluctuate}:
\begin{equation}
\label{int-N} \int dE'\,\varrho(E') =N.
\end{equation}
This normalization condition leads to
$$
\int
dE'\,\rho(E)\,\rho(E')\,R_{N}(E,E')=N\,\langle\varrho(E)\rangle=N\,\rho(E),
$$
which implies that
\begin{equation}
\label{SR} \int_{-\infty}^{+\infty} dE'\,\rho(E')\,Y_{N}(E,E')=-1,
\;\;\;\;\;\int_{-\infty}^{+\infty} dE'\,\rho(E')\,[R_{N}(E,E')-1]=0.
\end{equation}
This is the {\it normalization sum rule}.

Note that the sum rule can only be proven if $N$ is finite and
integration in Eq.(\ref{SR}) are extended over ${\it all}$ energies.
Taking the limit $N\rightarrow\infty$ could be a dangerous procedure
as in this case one has to worry about the commutativity  of the
limits $N\rightarrow\infty$ and {\it limits of
integration}$\rightarrow\infty$. The sum rule is certainly satisfied
if the limit $N\rightarrow\infty$ is done {\it after} doing the
integral. A simple example below shows that it can be violated if
the limit $N\rightarrow$ is taken {\it prior} of doing the integral.

Consider an ensemble of diagonal random matrices with independently
fluctuating random elements each having a distribution
\begin{equation}
\label{ddist} P(\varepsilon_{n})=\left\{ \begin{matrix} N^{-1}, &
|\varepsilon_{n}|<N/2\cr 0 &
|\varepsilon_{n}|>N/2\cr\end{matrix}\right.
\end{equation}
The TLCF for this ensemble can be computed straightforwardly:
$$
\langle\varrho(E)\,\varrho(E') \rangle=\sum_{n\neq
m}\int_{-N/2}^{N/2}\frac{d\varepsilon_{n}}{N}\int_{-N/2}^{N/2}\frac{d\varepsilon_{m}}{N}\,
\delta(E-\varepsilon_{n})\,\delta(E'-\varepsilon_{m})+\sum_{n}\int_{-N/2}^{N/2}\frac{d\varepsilon_{n}}{N}\,
\delta(E-\varepsilon_{n})\,\delta(E'-\varepsilon_{n}),
$$
so that for $|E|<N/2$ and $|E'|<N/2$ one obtains:
\begin{equation}
\label{diag-TLCF}
R(E,E')=\delta(E-E')+\frac{N^{2}-N}{N^{2}},\;\;\;\;\;\rho(E)=1,\;\;\;\;Y_{N}(E,E')=-\frac{1}{N}.
\end{equation}
At a finite $N$ the normalization sum rule Eq.(\ref{SR}) is
obviously fulfilled:
\begin{equation}
\label{SR-d} \int_{-N/2}^{N/2}dE'\,\left(-\frac{1}{N} \right)=-1.
\end{equation}
However, if one takes the limit $N\rightarrow\infty$ in
Eq.(\ref{diag-TLCF}) before integrating, one obtains $Y_{\infty}=0$,
and the normalization sum rule will we violated:
\begin{equation}
\label{SR-d}
\int_{-N/2}^{N/2}dE'\,\lim_{N\rightarrow\infty}Y_{N}(E,E')=0.
\end{equation}
This mechanism of violation of sum rules in the thermodynamic limit
is called the {\it anomaly} and is well known in the field theory.

A remarkable property of the Wigner-Dyson level statistics is that
in this case (as well for all invariant RM ensembles, even with
shallow confinement potentials) the normalization sum rule is not
violated and the anomaly does not occur. Let us show how this
property follows from the plasma model Eq.(\ref{densityFun}). To
this end we note that the density-density correlation function is
given by the variational derivative of the mean density with respect
to the confinement potential:
\begin{equation}
\label{var} \langle\varrho(E)\,\varrho(E')
\rangle-\langle\varrho(E)\rangle\,\langle\varrho(E')
\rangle=-\frac{\delta\rho(E)}{\delta V(E')}\approx
\rho_{0}^{2}\,[R_{\infty}(E-E')-1],
\end{equation}
where we assume that the mean density $\rho(E)\approx\rho_{0}$ does
not change much at a scale of the mean level spacing $\Delta$. In
this case one can approximately consider the TLCF as a function of
the energy difference. Then integrating by parts in Eq.(\ref{Muskh})
and neglecting the energy dependence of the square roots we obtain:
\begin{equation}
\label{TLCF-as} -\frac{\delta\rho(E)}{\delta
V(E')}=-\frac{1}{\pi^{2}\beta}\,\frac{1}{(E-E')^{2}},
\end{equation}
where the regularization
$$
(E-E')^{-2}\rightarrow \frac{1}{2}\,[(E-E'+i0)^{-2}+(E-E'-i0)^{-2}]
$$
is assumed. Using this regularization one can immediately check that
$$
\int_{-\infty}^{+\infty}[R_{\infty}(E)-1]\,dE=0,
$$
as it is required by the sum rule Eq.(\ref{SR}) at $\rho(E)={\rm
const}$. Thus we see that for the plasma model doing the limit
$N\rightarrow\infty$ and doing the integral commute.

 The absence of
the anomaly is related with the incompressible character of the
system of logarithmically repelling particles. This is the reason
why approximations made in deriving the plasma model
Eq.(\ref{densityFun}) did not affect the regular fulfillment of the
normalization sum rule. Below we define the {\it level
compressibility} and show that it is zero if the normalization sum
rule is not violated. To this end we define the {\it level number
variance}:
\begin{equation}
\label{lnv} \Sigma(\bar{n})=\langle n^{2}\rangle-\bar{n}^{2},
\end{equation}
where $n$ is the fluctuating number of levels in an energy interval
$\delta_{E}$ that contains on the average $\bar{n}$ levels. Writing
$$
n=\int_{E_{0}-\delta E/2}^{E_{0}+\delta
E/2}\rho(E)\,dE\equiv\int_{\delta_{E}}\rho(E)\,dE
$$
we obtain:
$$
\Sigma(\bar{n})=\int_{\delta_{E}} \rho(E)\,dE
\int_{\delta_{E}}\rho(E')\,[R_{N}(E,E')-1]\,dE'.
$$
Now we assume that $N$ is large, the confinement is steep and thus
$D\gg|\delta E|\gg \Delta$. This allows to consider
$\rho(E)=\rho_{0}$ and $R_{N}(E,E')\approx R_{\infty}(E-E')$. Then
one can integrate over $E$ at a fixed $E-E'$ and arrive at:
\begin{equation}
\label{Sig}
\Sigma(\bar{n})=\bar{n}\,\int_{-\bar{n}}^{\bar{n}}[R_{\infty}(s)-1]\,ds
-\int_{-\bar{n}}^{\bar{n}}|s|\,[R_{\infty}(s)-1]\,ds,
\end{equation}
where $s=(E-E')\rho_{0}$. The {\it level compressibility} is defined
as
\begin{equation}
\label{compress}
\chi(\bar{n})=\frac{d\Sigma}{d\bar{n}}=\int_{-\bar{n}}^{\bar{n}}[R_{\infty}(s)-1]\,ds.
\end{equation}
In the case where the anomaly does not occur we have
$$
\lim_{\bar{n}\rightarrow\infty}\int_{-\bar{n}}^{\bar{n}}[R_{\infty}(s)-1]\,ds=
\lim_{N\rightarrow\infty}\int_{-\infty}^{+\infty}[R_{N}(s)-1]\,ds,
$$
which in view of the normalization sum rule Eq.(\ref{SR}) implies
incompressible nature of the system of energy levels:
\begin{equation}
\label{chi0} \lim_{\bar{n}\rightarrow\infty}\chi(\bar{n})=0.
\end{equation}
The opposite is also true: if Eq.(\ref{chi0}) is fulfilled, the
limits ${\bar{n}\rightarrow\infty}$ and $N\rightarrow\infty$ commute
and there is no anomaly.

Now we use the absence of the anomaly and the normalization sum rule
to compute the level number variance at $\bar{n}\gg 1$. To this end
we cast Eq.(\ref{Sig}) in the following way
$$
\Sigma(\bar{n})=\bar{n}\,\int_{-\infty}^{+\infty}[R_{\infty}(s)-1]\,ds
-2\bar{n}\int_{\bar{n}}^{\infty}[R_{\infty}(s)-1]\,ds
-2\int_{0}^{\bar{n}}|s|\,[R_{\infty}(s)-1]\,ds,
$$
The first integral vanishes because of the sum rule Eq.(\ref{SR}),
the second term requires only the knowledge of TLCF at large
distances $s\gg 1$ and can be computed using Eq.(\ref{TLCF-as}) and
appears to be a constant of order 1. The third integral is
logarithmic and this allows to compute the leading logarithmic term
also using Eq.(\ref{TLCF-as}) which is valid at $s\gg 1$. Cutting
the logarithmic divergency at $s\sim 1$ we obtain for $\bar{n}\gg
1$:
\begin{equation}
\label{Sigma}
\Sigma(\bar{n})=\frac{2}{\pi^{2}\beta}\,\ln\bar{n}+O(1).
\end{equation}
This variance is considerably smaller than for independently
fluctuating levels (diagonal RME) where it is distributed according
to Poisson law:
\begin{equation}
\label{Poisson} \Sigma(\bar{n})=\bar{n}.
\end{equation}
\section{Orthogonal polynomials and energy level statistics for $\beta=2$.}
As has been already mentioned level statistics in the Gaussian
invariant RME with $\beta=2$ can be exactly mapped onto the system
of non-interacting fermions in one dimension in the parabolic
confinement potential $V(E)=E^{2}$. Let us check this statement. To
this end we recall that the ground state {\it many-body}
wavefunction of non-interacting fermions is the {\it Slatter
determinant}:
\begin{equation}
\label{Slatter}
\Psi(\{E_{n}\})=\left|\begin{matrix}\varphi_{0}(E_{1})&\varphi_{0}(E_{2})&...&
\varphi_{0}(E_{N})\cr \varphi_{1}(E_{1})&\varphi_{1}(E_{2})&...&
\varphi_{1}(E_{N})\cr...&...&...&...\cr
\varphi_{N-1}(E_{1})&\varphi_{N-1}(E_{2})&...&
\varphi_{N-1}(E_{N})\cr
\end{matrix} \right|.
\end{equation}
The {\it one-particle} eigenfunctions
$\varphi_{n}(E)=H_{n}(E)\,e^{-E^{2}/2}$ in the parabolic confinement
potential $V(E)$ obeying the Schroedinger equation:
\begin{equation}
\label{Schred} -\frac{1}{2}\,\frac{\partial^{2}}{\partial
E^{2}}\,\varphi_{n}(E)+\frac{1}{2}\,E^{2}\,\varphi_{n}(E)={\cal
E}_{n}\,\varphi_{n}(E),
\end{equation}
are related to the Hermite {\it orthogonal polynomials} $H_{n}(E)$.
These are the polynomials of the $n$-th order satisfying the
orthogonality relation:
\begin{equation}
\label{ortho} \int_{-\infty}^{+\infty}{\rm
exp}[-V(x)]\,H_{n}(x)\,H_{m}(x)=h_{n}\,\delta_{nm},\;\;\;\;\;\;V(x)=x^{2},\;\;\;\;h_{n}=1.
\end{equation}
It is important that the Hermite polynomials obey the {\it
three-term} recursive relation:
\begin{equation}
\label{three-term} H_{n+1}(x)=2x\,H_{n}(x)-2n\,H_{n-1}(x).
\end{equation}
Using this relation one can show that the Slatter determinant
Eq.(\ref{Slatter}) reduces to:
\begin{equation}
\label{Sl-VdM}
\left|\begin{matrix}\varphi_{0}(E_{1})&\varphi_{0}(E_{2})&...&
\varphi_{0}(E_{N})\cr \varphi_{1}(E_{1})&\varphi_{1}(E_{2})&...&
\varphi_{1}(E_{N})\cr...&...&...&...\cr
\varphi_{N-1}(E_{1})&\varphi_{N-1}(E_{2})&...&
\varphi_{N-1}(E_{N})\cr
\end{matrix} \right|={\rm const}\,
\left|\begin{matrix}1&1&...&1\cr E_{1}&E_{2}&...&
E_{N}\cr...&...&...&...\cr E_{1}^{N-1}&E_{2}^{N-1}&...&
E_{N}^{N-1}\cr
\end{matrix} \right|\,\rm exp\left[-\frac{1}{2}\,\sum_{n=1}^{N} E_{n}^{2}\right].
\end{equation}
Indeed, the exponential factors $e^{-\frac{1}{2}E^{2}}$ in all the
$\varphi_{n}(E)$ can be taken out of the determinant using the rule
of multiplication of  determinant by a factor which is equivalent to
multiplication of all the elements in a column by this factor. So we
obtain the exponential factor in the r.h.s. of Eq.(\ref{Sl-VdM}).

Next choosing $H_{0}(E)=1$, $H_{1}(E)=x$ one can find all the other
polynomials using the recursion relation Eq.(\ref{three-term}). In
particular, $H_{2}=2x^{2}-2$. This polynomial should be plugged in
the third line of the determinant in the l.h.s. of
Eq.(\ref{Sl-VdM}). Note, however, that the constant term $-2$ can be
omitted as its inclusion is equivalent to an addition of the first
line to the third line in the determinant which according to the
basic property of a determinant does not change its value. This
process can be continued. For instance in $H_{3}=2x H_{2}-4H_{1}$
which stands in the fourth line of the determinant one can omit
$-4H_{1}$, as $H_{1}$ stands in the second line. Then expressing
$H_{2}=2x H_{1}-H_{0}=2x^{2}-1$  one can put in the fourth line
$2x(2x^{2} -1)$ instead of $H_{3}$. Finally, observing that the term
$-2x$ can be considered as a linear combination of $4x^{3}$ and the
first line in the determinant and omitting this term we conclude
that instead of $H_{3}$ on can put in the fourth line of the
determinant just one term $4x^{3}$.

The determinant in the r.h.s. is the famous {\it Vandermond
determinant} which is equal to:
\begin{equation}
\label{VdM} \left|\begin{matrix}1&1&...&1\cr E_{1}&E_{2}&...&
E_{N}\cr...&...&...&...\cr E_{1}^{N-1}&E_{2}^{N-1}&...&
E_{N}^{N-1}\cr
\end{matrix} \right|= \prod_{n>m}(E_{n}-E_{m}).
\end{equation}
Now we see that:
\begin{equation}
\label{psi-2} |\Psi(\{ E_{n}\})|^{2}={\rm const}\,{\rm
exp}\left[-\sum_{n}E_{n}^{2}\right]\,\prod_{n>m}(E_{n}-E_{m})^{2}.
\end{equation}
This is exactly the probability distribution functions for the
eigenvalues of the Gaussian RME with $\beta=2$.

It turns out that the theory of orthogonal polynomials \cite{Sciego}
is the powerful method to solve {\it any} orthogonal random matrix
ensemble. What one has to do for that is to generate a set of
orthogonal polynomials $p_{n}(x)$ obeying the orthogonality relation
Eq.(\ref{ortho}) and to be able to compute the large-$N$ asymptotic
behavior of the "wavefunctions":
\begin{equation}
\label{varphi-n}
\varphi_{n}(E)=\varphi_{n}(E)=h_{n}^{-1/2}\,p_{n}(E)\,e^{-V(E)/2}.
\end{equation}
The generation of orthogonal polynomials is possible for any
confinement potential $V(x)$ using the Gram-Schmidt
orthogonalization procedure. According to this procedure one
computes the Gram-Schmidt determinant:
\begin{equation}
\label{GSdet} G_{n}= \left|\begin{matrix}a_{0}&a_{1}&...&a_{n}\cr
a_{1}&a_{2}&...& a_{n+1}\cr...&...&...&...\cr a_{n}&a_{n+1}&...&
a_{2n}\cr
\end{matrix} \right|,
\end{equation}
where $a_{n}$ are the moments:
\begin{equation}
\label{mom} a_{n}=\int_{-\infty}^{+\infty}{\rm
exp}[-V(x)]\,x^{n}\,dx.
\end{equation}
Then the orthogonal polynomial of $n$-th power $p_{n}$ is given by:
\begin{equation}
\label{orpol}
p_{n}(x)=\frac{q_{n}}{G_{n-1}}\,\left|\begin{matrix}a_{0}&a_{1}&...&a_{n}\cr
a_{1}&a_{2}&...& a_{n+1}\cr...&...&...&...\cr a_{n-1}&a_{n}&...&
a_{2n-1}\cr1&x&...&x^{n}
\end{matrix} \right|,
\end{equation}
where $q_{n}$ is the coefficient in front of $x^{n}$ in $p_{n}$.

It follows from this generic procedure that any set of orthogonal
polynomials $p_{n}(x)$ should obey the three-term recursion relation
similar to Eq.(\ref{three-term}). For the confinement potential
$V(x)$ being an even function of $x$ and  the choice $q_{n}=1$ (for
the Hermite polynomials the standard definition corresponds to
$q_{n}=2^{n}$) it reads:
\begin{equation}
\label{gen-3-term}
p_{n+1}(x)=xp_{n}(x)-C_{n+1}\,p_{n-1}(x),\;\;\;\;\;\;C_{n+1}=\frac{G_{n}\,G_{n-2}}{G_{n-1}^{2}},
\;\;\;p_{0}(x)=1,\;\;\;\;p_{1}(x)=x.
\end{equation}
We note that this recursion relation generates orthogonal but not
{\it ortho-normal} polynomials. The price of having $q_{n}=1$ is
that the normalization constant $h_{n}$ in Eq.(\ref{ortho}) is not
unity and is related to the coefficient $C_{n}$ as follows:
\begin{equation}
\label{h-C} h_{n}=h_{0}\,\prod_{m=2}^{n+1}C_{m},\;\;\;\;h_{0}=a_{0}.
\end{equation}
The recursive relation Eq.(\ref{gen-3-term}) appears to be the most
convenient way of generating orthogonal polynomials for {\it any}
confinement potential. It works also for non-classical polynomials
for which there are no second-order differential equations (similar
to the Schroedinger equation Eq.(\ref{Schred}) in the case of
Hermite polynomials) which the "wavefunctions" Eq.(\ref{varphi-n})
should obey.

The efficiency of the orthogonal polynomials in the problem of level
statistics is largely due to the {\it Christoffel-Darboux formula}:
\begin{equation}
\label{Chr-Dar}
K_{N}(x,y)=\sum_{n=0}^{N-1}\varphi_{n}(x)\,\varphi_{n}(y)=\sqrt{\frac{h_{N}}{h_{N-1}}}\,\,
\frac{\varphi_{N-1}(x)\,\varphi_{N}(y)-\varphi_{N-1}(y)\,\varphi_{N}(x)}{y-x}.
\end{equation}
This formula can be proven by induction using the three term
recursive relation Eq.(\ref{gen-3-term}) and a relation
Eq.(\ref{h-C}) between $h_{n}$ and $C_{n}$.

The Christoffel-Darboux formula  is important because the mean
density of states and the two level correlation function at $x\neq
y$ are given by:
\begin{equation}
\label{mDoS} \rho(E)=\sum_{n=0}^{N-1}\varphi_{n}(x)^{2}=K_{N}(E,E),
\end{equation}
\begin{equation}
\label{TLCF-K} \rho(E)\,\rho(E')\,R_{N}(E,E')=
\sum_{n=0}^{N-1}\sum_{m=0}^{N-1}\left[\varphi^{2}_{n}(E)\,\varphi^{2}_{m}(E')-
\varphi_{n}(x)\,\varphi_{n}(y)\,\varphi_{m}(y)\,
\varphi_{m}(x)\right]= K_{N}(E,E)\,K_{N}(E',E')-K^{2}_{N}(E,E').
\end{equation}
Eqs.(\ref{mDoS}),(\ref{TLCF-K}) can be proven formally without any
reference to systems of non-interacting fermions. However, it is
instructive to see how they follow from the fermionic second
quantization formalism. Indeed, the density of non-interacting
fermions and the density-density correlation function are given by:
$$
\langle 0|\hat{\Psi}^{\dagger}(x)\hat{\Psi}(x)|0\rangle,\;\;\;\;\;
\langle
0|\hat{\Psi}^{\dagger}(x)\hat{\Psi}(x)\,\hat{\Psi}^{\dagger}(y)\hat{\Psi}(y)|0\rangle,
$$
where $\langle 0|...|0\rangle$ is the quantum-mechanical averaging
over the ground state. According to the rules of second quantization
the fermionic field operator $\hat{\Psi}(x)$ is given by the
expansion over the single-particle wavefunctions:
$$
\hat{\Psi}(x)=\sum_{n=0}^{N-1}\varphi_{n}(x)\,c_{n},
$$
where $c_{n}$ and $c_{n}^{\dagger}$ are the fermionic creation and
annihilation operators obeying the anti-commutation relation
$$
c_{n}^{\dagger}\,c_{m}+c_{m}\,c_{n}^{\dagger}=\delta_{nm}.
$$
The averages over the ground state can be computed using the Wick
theorem:
\begin{equation}
\label{Wick} \langle0|
c_{n}^{\dagger}c_{m}|0\rangle=\delta_{nm},\;\;\;\;\;\;\langle0|
\,c_{n_{1}}^{\dagger}c_{n_{2}}
c_{n_{3}}^{\dagger}c_{n_{4}}|0\rangle=\delta_{n_{1}n_{2}}\,\delta_{n_{3}n_{4}}-
\delta_{n_{1}n_{4}}\,\delta_{n_{2}n_{3}}.
\end{equation}
The two terms in Eq.(\ref{TLCF-K}) follow from the two terms in
Eq.(\ref{Wick}) which correspond to two possible parings of
$c^{\dagger}$ and $c$.

Eq.(\ref{TLCF-K}) can be conveniently represented in terms of the
determinant:
\begin{equation}
\label{TLCF-det}
\rho(E)\,\rho(E')\,R_{N}(E,E')=\left|\begin{matrix}K_{N}(E,E)&
K_{N}(E,E')\cr K_{N}(E',E)&K_{N}(E',E')\cr
\end{matrix} \right|.
\end{equation}
Using the fermionic analogy and the Wick theorem  one can prove that
any multi-point level density correlation function can be
represented in the form of a similar determinant:
\begin{equation}
\label{multi}
\rho(E_{1})\rho(E_{2})...\rho(E_{n})\,R(E_{1},E_{2}...E_{n})=\left|\begin{matrix}
K_{N}(E_{1},E_{1})&K_{N}(E_{1},E_{2})&...&K_{N}(E_{1},E_{n})\cr
K_{N}(E_{2},E_{1})&K_{N}(E_{2},E_{2})&...&
K_{N}(E_{2},E_{n})\cr...&...&...&...\cr
K_{N}(E_{n},E_{1})&K_{N}(E_{n},E_{2})&...& K_{N}(E_{n},E_{n})\cr
\end{matrix} \right|.
\end{equation}
We see that the analogy with non-interacting fermions allows to
express any multi-point level density correlation function in terms
of only one single kernel $K_{N}(x,y)$. The latter according to the
Christoffel-Darboux theorem is a product of only two "wave
functions" which require the knowledge of only  two orthogonal
polynomials $p_{N}(x)$ and $p_{N-1}(x)$. Thus the problem of energy
level statistics is reduced to the problem of finding the asymptotic
behavior of orthogonal polynomials of high order.
\section{WKB quasi-classical approximation and the one-dimensional "Wigner crystal".}
The semicircle law for the mean level density follows immediately
and trivially from the free-fermion representation. Indeed, the
density of one-dimensional fermions is directly related with the
Fermi-momentum $p_{F}$ by:
\begin{equation}
\label{Fermi-np} \frac{2p_{F}}{2\pi}=\rho,\;\;\;\;\;\hbar=1.
\end{equation}
When the density varies slowly at a scale of the Fermi wavelength,
one can apply Eq.(\ref{Fermi-np}) {\it locally} thus relating the
local Fermi-momentum $p_{F}(x)$ with the local density $\rho(x)$.
The local Fermi momentum corresponds to the momentum of the highest
occupied state in a parabolic potential
\begin{equation}
\label{local pF} p_{F}=p_{N}(x)=\sqrt{2m E_{\rm kin}}=\sqrt{2{\cal
E}_{n}-x^{2}}=\sqrt{2N+1 -x^{2}},\;\;\;\;\;m=1.
\end{equation}
Then the local density is obtained immediately from
Eq.(\ref{Fermi-np}):
\begin{equation}
\label{semicircle2}
\rho(x)=\frac{1}{\pi}\,p_{N}(x)=\frac{1}{\pi}\,\sqrt{2N+1 -x^{2}}.
\end{equation}
This is the celebrated {\it semicircle law}.

In order to obtain the {\it density-density correlation function},
or the {\it two-level correlation function} one should work a little
bit harder determining the large $N$ asymptotic behavior of
$\varphi_{n}$ and applying Eq.(\ref{TLCF-K}).

In the case of Hermite polynomials the problem of the large-$N$
asymptotic behavior  can be solved quite easily. The reason is that
there is the second-order differential equation (the Schroedinger
equation) which the wavefunctions $\varphi_{N}(x)$ must obey. As is
well known the solutions to the Schroediger equation corresponding
to large quantum numbers bear the properties of classical motion in
the corresponding potential. In quantum mechanics this corresponds
to the "quasi-classical", or WKB approximation \cite{LL}.

According to this approximation the wave function at $x^{2}< 2N$ is
proportional to:
\begin{equation}
\label{WKB}
\varphi_{N}(x)\propto[p_{N}(x)]^{-1/2}\,\left\{\begin{matrix}\cos\left[\int_{0}^{x}
p_{N}(x')\,dx'\right]& N&is&even\cr& & & \cr\sin\left[\int_{0}^{x}
p_{N}(x')\,dx'\right]&N&is&odd\cr
\end{matrix}\right.
\end{equation}
At large $N$ we obtain:
$$
p_{N}(x)\approx \sqrt{2N},\;\;\;\;\;\int_{0}^{x}
p_{N}(x')\,dx'\approx x\,\sqrt{2N}.
$$
Then the kernel $K_{N}(x,y)$ is easily calculated using the
Christoffel-Darboux formula Eq.(\ref{Chr-Dar}):
$$
\label{kernEL} K_{N}(x,y)={\rm
const}\,\frac{\sin\left(\sqrt{2N}\,(x-y)\right)}{x-y},\;\;\;\;\;K_{N}(x,x)={\rm
const}\,\sqrt{2N}.
$$
The normalization constant  ${\rm const}=1/\pi$ is most easily found
from the comparison of $K_{N}(x,x)=\rho(x)$ and the semi-circle mean
level density Eq.(\ref{semi-circ}) which at large $N$ reduces to
$\rho(x)\approx \sqrt{2N}/\pi=\rho_{0}$. Now, introducing mean level
spacing $\Delta=\rho^{-1}\approx \pi/\sqrt{2N}$ we arrive at:
\begin{equation}
\label{kernEL} K_{N}(x,y)\rightarrow
K(s)=\rho_{0}\,\frac{\sin\left(\pi\,s\right)}{\pi
s},\;\;\;\;\;s=\frac{x-y}{\Delta},\;\;\;\;\;\;\;N\rightarrow\infty.
\end{equation}
The two level correlation function Eq.(\ref{TLCF-K}) is then equal
to:
\begin{equation}
\label{TLCF-beta2}
R_{\infty}(x,y)=\delta(s)+1-\frac{\sin^{2}\left(\pi\,s\right)}{(\pi
s)^{2}}=1-\frac{1}{2\pi^{2}\,s^{2}}+\frac{\cos(2\pi\,
s)}{2\pi^{2}\,s^{2}}.
\end{equation}
One can see that the TLCF given by Eq.(\ref{TLCF-beta2}) has all the
asymptotic limits right. It is proportional to $s^{2}$ at $s\ll 1$
and its envelope corresponds to Eq.(\ref{TLCF-as}) for $s\gg 1$. In
addition to that it obeys the normalization sum rule Eq.(\ref{SR}).
However, in Eq.(\ref{TLCF-beta2}) there is a term that oscillates
with the period of the mean level spacing $\Delta$. This term evades
the consideration based  on the $2\times2$ matrix (small $s\ll 1$)
and the effective continuous plasma model (large $s\gg 1$).

Let us discuss the physical meaning of this term using the plasma
model analogy but without the continuous approximation. It is well
known that plasma of particles with the long-range repulsion in a
confinement potential tend to develop a crystal order known as
Wigner crystal. Such Wigner crystal of electrons have been observed
on top of the helium surface. Our case is special, as it is
one-dimensional. According to the Mermin theorem the crystal order
cannot survive in one dimensions at a finite temperature because
thermal fluctuations destroy the long-range order. However, local
crystal order may exist. The last oscillating them in
Eq.(\ref{TLCF-beta2}) reflects exactly this order. The short-range
nature of this order manifests itself in the fast $s^{-2}$ decay of
oscillations at large distances.

So far in this section we have considered the $\beta=2$ case. As in
the plasma analogy $\beta$ plays a role of inverse temperature, one
would expect the oscillating term to decay slower for $\beta=4$ and
faster for $\beta=1$.   This expectation is in fact true.

One can show using the more sophisticated application of the
orthogonal polynomial machinery that in the limit
$N\rightarrow\infty$ the two-level correlation functions for the
orthogonal ($\beta=1$) ensemble $R_{\infty}^{{\rm orth}}$ and that
for the symplectic ($\beta=4$) ensemble  $R_{\infty}^{{\rm symp}}$
can also be expressed in terms of the kernel $K_{N}(s)$,
Eq.(\ref{kernEL}):
\begin{equation}
\label{orth-R} Y_{\infty}^{{\rm
orth}}(s)=-K^{2}(s)-\frac{dK(s)}{ds}\,\int_{s}^{\infty}K(x)\,dx,
\end{equation}
\begin{equation}
\label{symp-R} Y_{\infty}^{{\rm
symp}}(s/2)=-K^{2}(s)+\frac{dK(s)}{ds}\,\int_{0}^{s}K(x)\,dx.
\end{equation}
The $s/2$ in the argument of $Y_{\infty}^{{\rm symp}}(s/2)$ appears
because of the Kramers degeneracy: for the same total number of
levels $N$ the mean level spacing between doubly degenerate levels
is two times longer. Accordingly, the $\delta(s)$ function in $R(s)$
enters with the pre-factor of 2. The asymptotic behavior of these
functions for $s\gg1$ is the following:
\begin{equation}
\label{orth-asy} Y_{\infty}^{{\rm
orth}}(s)=-\frac{1}{\pi^{2}\,s^{2}}+\frac{\cos(2\pi
s)}{2\pi^{4}s^{4}},
\end{equation}
\begin{equation}
\label{symp-asy} Y_{\infty}^{{\rm
symp}}(s)=-\frac{1}{4\pi^{2}\,s^{2}}+\frac{\cos(2\pi\,
s)}{4\,s}+\frac{\cos(4\pi\,s)}{2(2\pi\,s)^{4}}.
\end{equation}
One can see that the leading oscillating term decrease as
$s^{-4/\beta}$ as was expected. In addition to that, in the
symplectic ensemble $\beta=4$, the sub-leading second-harmonic term
appears which was absent to all orders in $1/s$ for $\beta=1,2$.

The spectral correlations of a quantum system show up in the
time-dependence of response to external time-dependent
perturbations. For such applications one need to know the
Fourier-transform $F(t)$ of the two-level correlation (cluster)
function $Y_{\infty}(s)$. It appears to be amazingly simple for the
unitary ensemble $\beta=2$:
\begin{equation}
\label{Funi} F^{{\rm unit}}=\left\{\begin{matrix}|t|-1,&|t|<1\cr
0,&|t|>1.
\end{matrix}\right.
\end{equation}
with the jump of the first derivative at $t=1$ that leads to the
oscillations with the period 1 which amplitude decreases as
$1/s^{2}$. For the orthogonal ensemble $\beta=1$ there is a jump
only in the third derivative:
\begin{equation}
\label{Fortho} F^{{\rm
ortho}}=\left\{\begin{matrix}2|t|-1-|t|\,\ln(1+2|t|),&|t|<1\cr
1-|t|\,\ln\left( \frac{2|t|+1}{2|t|-1}\right),&|t|>1.
\end{matrix}\right.
\end{equation}
For the symplectic ensemble $\beta=4$ there are {\it two} singular
points: $|t|=1$ and $|t|=2$ which correspond to two oscillating
terms in Eq.(\ref{symp-asy}):
\begin{equation}
\label{Fortho} F^{{\rm
symp}}=\left\{\begin{matrix}\frac{1}{2}\,|t|-1-\frac{1}{4}\,|t|\,\ln|1-|t||,&|t|<2\cr
0,&|t|>2.
\end{matrix}\right.
\end{equation}
\section{Wigner-Dyson level statistics and the Luttinger liquid.}
The large-$s$ asymptotics  of the two-level correlation function
containing both the non-oscillating and the oscillating terms which
decay as a certain power-law can be written in a compact form which
involves only one single function $G(s)$:
\begin{equation}
\label{AA-u} Y^{{\rm
unit}}_{\infty}(s)=-\frac{1}{4\pi^{2}}\,\frac{\partial^{2}G(s)}{\partial
s^{2}}+\cos(2\pi s)\,e^{G(s)},
\end{equation}
\begin{equation}
\label{AA-o} Y^{{\rm
ortho}}_{\infty}(s)=-\frac{1}{2\pi^{2}}\,\frac{\partial^{2}G(s)}{\partial
s^{2}}+2\cos(2\pi s)\,e^{2G(s)},
\end{equation}
\begin{equation}
\label{AA-s} Y^{{\rm
symp}}_{\infty}(s)=-\frac{1}{8\pi^{2}}\,\frac{\partial^{2}G(s)}{\partial
s^{2}}+\frac{\pi}{\sqrt{8}}\,\cos(2\pi
s)\,e^{G(s)/2}+\frac{1}{8}\,\cos(4\pi s)\,e^{2G(s)},
\end{equation}
where
\begin{equation}
\label{G} G=-\ln(2\pi^{2}s^{2}).
\end{equation}
It turns out that the function $G(x)$ is proportional to the
equal-time correlation function of a free bosonic field
$\Phi(x,\tau)$ in the {\it two} dimensional space-(imaginary)time,
which arises as a bosonized version of the Calogero-Sutherland model
Eq.(\ref{CSM}) of interacted fermions. More generally, a great
number of models of interacted electrons in one dimension fall into
the universality class of {\it Luttinger liquid}\cite{Shurabook}
which is characterized by a certain correlation functions at {\it
large} separations in space and/or in time. All of them follow from
the fact that the fermionic operator $\Psi(x,\tau)$ can be
represented as
\begin{equation}
\label{psi-phi}
\Psi(x,\tau)=R\,e^{ik_{F}x}+Le^{-ik_{F}x},\;\;\;\;\;R,\,L=\frac{1}{\sqrt{\pi}}\,{\rm
exp}[\pm i\Phi_{R,L}(x,\tau)],
\end{equation}
where $
\Phi(x,\tau)=\frac{1}{2}\,[\Phi_{R}(x,\tau)+\Phi_{L}(x,\tau)]$ is
the {\it free boson field} with the action:
\begin{equation}
\label{bos-act} S[\Phi]=\frac{1}{2\pi
K}\int_{0}^{1/T}d\tau\,\int_{-\infty}^{+\infty}dx\,[(\partial_{x}\Phi)^{2}+
(\partial_{\tau}\Phi)^{2}]=\frac{L}{\pi K
T}\sum_{k>0,\omega_{n}=2\pi T
n}|A_{k,\omega_{n}}|^{2}\,(\omega_{n}^{2}+k^{2}),
\end{equation}
where
\begin{equation}
\label{kw} \Phi(x,\tau)=\sum_{k>0,\omega_{n}=2\pi T\,n}\left\{
A_{k,\omega}e^{i(\omega_{n}\tau+kx)}+c.c\right\},
\end{equation}
$\omega_{n}=2\pi T n$ ($n$ are {\it all} integers) and $k=(2\pi/L)m
>0$ to avoid double-counting. The action Eq.(\ref{bos-act})
corresponds to:
\begin{equation}
\label{var} \langle|A_{k,\omega}|^{2}\rangle = \frac{\pi K
T}{L}\,\frac{1}{k^{2}+\omega_{n}^{2}}.
\end{equation}
It is remarkable that {\it interaction} of fermions is encoded in
only one single parameter $K$ which is $K<1$ for repulsion, $K=1$
for the non-interacting fermions and $K>1$ for attraction. In other
words, with respect to long-distance properties the system of
interacted fermions in one dimensions ($1+1$ space-time) is
equivalent to a system of free bosons. The physical meaning of this
result  is that for systems of the Luttinger-liquid universality
class all the multitude of effects of electron interaction reduces
to dynamics and thermodynamics of the plasmon collective modes.

The density operator in this representation is given by:
\begin{equation}
\label{dens-LL}
\rho(x,\tau)-\rho_{0}=\frac{1}{\pi}\,\partial_{x}\Phi(x,\tau)+A_{1}\,\cos[2k_{F}x+
2\Phi(x,\tau)]+A_{2}\,\cos[4k_{F}x+4\Phi(x,\tau)]+...,
\end{equation}
where $A_{k}$ are the structural constants which are determined from
the details of the system at small distances.

The first term in Eq.(\ref{dens-LL}) comes from the combination
$[R^{+}R+L^{+}L]$ in $\Psi^{\dagger}\Psi$. It is analogous to the
$\delta\rho\propto \nabla{\bf u}$ term in hydrodynamics, where ${\bf
u}(x)$ is the mass displacement at a point $x$.   The correct
evaluation of this contribution requires the regularization
$R^{+}R=R^{+}(x+a,\tau)R(x,\tau)$ (and the similar regularization
for $L$), where $a=1$ is the lattice constant which corresponds to
$k_{F}=\pi$. Oscillating terms proportional to $e^{\pm 2ik_{F}x}$
arise from the cross combinations $R^{+}L+L^{+}R$ in
$\Psi^{\dagger}\Psi$. Note that for interacting fermion system there
is a vertex correction that involves momentum transfer far away from
the Fermi points $\pm k_{F}$. As the result, the density operator
expressed in terms of the field $\Psi(x,\tau)$ (which contains only
momenta close to the Fermi points) is not simply equal to
$\rho=\Psi^{\dagger}\Psi$ but may have higher order terms as well:
\begin{equation}
\label{high-or} \rho(x,\tau)=\Psi^{\dagger}\Psi+
c_{1}\,(\Psi^{\dagger}\Psi)^{2}+...
\end{equation}
It is these higher order terms that generate combinations containing
higher harmonics like $e^{\pm 4ik_{F}x}$.

Using Eq.(\ref{dens-LL}) and Eq.(\ref{bos-act}) one can express the
density-density correlation function through the free bosonic
correlation function. To this end we use the identity valid for any
Gaussian field theory:
$$
 \langle e^{i n\Phi(x,\tau)}\,e^{-i n\Phi(0,0)}\rangle={\rm
 exp}\left\langle
 -\frac{n^{2}}{2}\,(\Phi(x,\tau)-\Phi(0,0))^{2}\right\rangle_{S}.
$$
Now observe that for $s\gg 1$ and $T\rightarrow 0$ we have in the
thermodynamic limit $L\rightarrow\infty$:
$$
\langle \Phi(s,0)\Phi(0,0)\rangle_{S}-\langle
\Phi(0,0)\Phi(0,0)\rangle_{S}=-\frac{K}{2}\int_{0}^{\pi}dq\,\frac{1-\cos(qs)}
{q}\approx \frac{K}{4}\,G(s),
$$
where $G(s)$ is given by Eq.(\ref{G}). Finally, for oscillating part
of the density-density correlator we obtain:
\begin{equation}
\label{osc-ccor}
 \langle e^{i n\Phi(s,\tau)}\,e^{-i n\Phi(0,0)}\rangle={\rm
 exp}\left\{\frac{1}{4}K\,n^{2}\,G(s) \right\}.
\end{equation}
The non-oscillating part is expressed through the second derivative
of the correlation function of free boson field:
\begin{equation}
\label{nosc-ccor} \frac{1}{\pi^{2}}\,\langle
\partial_{s}\Phi(s,0)\partial_{s'}\Phi(s',0)\rangle_{S}=-\frac{K}{4\pi^{2}}\,\partial^{2}_{s}\,G(s).
\end{equation}
Eqs.(\ref{osc-ccor}),(\ref{nosc-ccor}) show that the phenomenology
of the Luttinger liquid allows to relate the coefficient in front of
the non-oscillating part of the density-density correlator with the
coefficients in front of $G(s)$ in the exponent determining the
amplitude of the oscillating terms. First of all we fix the
interaction parameter $K$ from the amplitude $1/(2\pi^{2}\beta)$ of
the non-oscillating part. Eq.(\ref{nosc-ccor}) suggests that:
\begin{equation}
\label{K-inter} K=\frac{2}{\beta}.
\end{equation}
Then a comparison of Eq.(\ref{osc-ccor}) with the oscillating terms
in Eqs.(\ref{AA-u}),(\ref{AA-o}),(\ref{AA-s}) shows that all
coefficients $\kappa$ in the amplitude of $n$-th harmonic
$e^{\kappa\,G(s)}\,\cos(2\pi n\, x)$ are equal to $\kappa=n^{2}K$
with $K$ given by Eq.(\ref{K-inter}). This is exactly the parameter
$K$ that corresponds to the Calogero-Sutherland model
Eq.(\ref{CSM}). Thus we have demonstrated that the Wigner-Dyson
level statistics at large level separations corresponds to the
particle statistics of the Calogero-Sutherland model at zero
temperature.

One can ask a question: how the Wigner-Dyson ensemble should be
deformed in order to retain this analogy with the
Calogero-Sutherland model also for finite temperatures $T\neq 0$.
The answer is \cite{KTs} that the proper deformation is given by the
Gaussian non-invariant ensemble Eq.(\ref{crit}) at large values of
the parameter $B$. The corresponding temperature of the
Calogero-Sutherland model is \cite{KTs}:
\begin{equation}
\label{T-B} T=\frac{1}{4\beta B}.
\end{equation}
The asymptotics of the correlation functions is given by
Eq.(\ref{AA-u})-(\ref{AA-s}) where one should substitute the
deformed function $G(s)$ in a compactified space-time rolled into a
cylinder of the circumference $1/T$ in the $\tau$ -direction:
\begin{equation}
\label{cyl} G_{T}(s)=-\ln\left( 2\pi^{2}\,\frac{\sinh^{2}(\pi T
s)}{(\pi T)^{2}}\right).
\end{equation}
This function is proportional to the Green's function of the free
bosonic field Eq.(\ref{bos-act}) at a finite temperature $T$ and can
be obtained from Eq.(\ref{var}) by {\it summing} over Matsubara
frequencies $\omega_{n}=2\pi T\,n$ instead of integrating over
$\omega$.

It is seen from Eqs.(\ref{AA-u})-(\ref{AA-s}) that breaking the
basis invariance by introducing the finite bandwidth $B$ and the
corresponding temperature $T$ of the bosonic system has an effect of
making the amplitudes of the oscillating terms exponentially
decaying for $s\gg 1/(\pi T)=(4\beta\,B)/\pi$.

\section{Field theories for random matrix ensembles}
In this section we derive the field theory for an arbitrary Gaussian
random matrix ensemble. This formalism, known as nonlinear
super-symmetric sigma-model \cite{Efet} has been first applied to
the Wigner-Dyson random matrix ensemble. However, its real strength
is in the possibility of extension to the non-invariant random
matrix ensembles as well as to real disordered conductors with
diffusive dynamics of particles.

We start by writing down the expression for retarded ($G^{R}$) or
advanced $G^{A}$ Green's function in terms of the functional
integral over complex variables $\varphi_{n}$:
\begin{equation}
\label{GF} G_{nm}^{R/A}=\left([E_{\pm}-{\bf
H}]^{-1}\right)_{nm}=\frac{\mp i}{Z}\,\int {\cal D}\varphi\,{\cal
D}\varphi^{*}\,\,\varphi_{n}\,\varphi_{m}^{*}\,{\rm exp}\left[
iS_{\pm}[\varphi]\right],
\end{equation}
where
\begin{equation}
\label{Spm}
S_{\pm}[\varphi]=\pm\sum_{i,j}\varphi_{i}^{*}\,\left[E_{\pm}\delta_{ij}-H_{ij}
\right]\,\varphi_{j}
\end{equation}
and $E_{\pm}=E\pm(\omega/2 + i 0)$. The sign $\pm$ stands for
$G^{R}$ or $G^{A}$, respectively. There would be no problem to
average Eq.(\ref{GF}) over the Gaussian random entries $H_{nm}$ if
not the normalization constant (partition function) $Z$:
\begin{equation}
\label{Z} Z=\int {\cal D}\varphi\,{\cal D}\varphi^{*}\,{\rm
exp}\left[ iS_{\pm}[\varphi]\right].
\end{equation}
which also depends on $H_{nm}$. With the $Z$ present one has a
problem, the {\it problem of denominator}.

There are different ways of overcoming this problem, e.g. the {\it
replica trick}. However, here we use another trick, the {\it
super-symmetry} method \cite{Efet}. In the core of this method is
the calculus of anti-commuting (Grassmann) variables $\mu_{m}$:
\begin{equation}
\label{mu} \mu_{n}\,\mu_{m}=-\mu_{m}\,\mu_{n},\;\;\;\;\mu_{n}^{2}=0.
\end{equation}
One can define the Grassmann integral with the shortest table of
integrals ever:
\begin{equation}
\label{int} \int
\mu_{n}\,d\mu_{n}=\frac{1}{\sqrt{\pi}},\;\;\;\;\;\int d\mu_{n}=0.
\end{equation}
Since any function $f(\mu_{n})=f(0)+f'(0)\,\mu_{n}$, the Grassmann
integration is essentially a differentiation. Now let us compute the
integral
\begin{equation}
\label{int-Gr} \int \,\prod_{i}d\mu^{*}_{i}d\mu_{i}\, {\rm
exp}\left[-\sum_{n}\mu_{n}^{*}\,a_{n}\,\mu_{n}
\right]=\int\,\prod_{i}d\mu^{*}_{i}d\mu_{i}\,\mu_{i}^{*}\mu_{i}\,(-a_{i})=
\int\,\prod_{i}\mu_{i}^{*}d\mu^{*}_{i}\,\mu_{i}
d\mu_{i}\,a_{i}=\prod_{i}\frac{a_{i}}{\pi}.
\end{equation}
To accomplish this we expand the exponential function to leave only
the term that contains a {\it complete} set
$\mu_{1}^{*}\mu_{1}...\mu_{n}^{*}\mu_{n}$ of Grassmann variables and
apply the table of integration Eq.(\ref{int}).

The corresponding integral over the usual complex variables would
give the following result:
\begin{equation}
\label{int-com} \int \,\prod_{i}d\varphi^{*}_{i}d\varphi_{i}\, {\rm
exp}\left[-\sum_{n}\varphi_{n}^{*}\,a_{n}\,\varphi_{n}
\right]=\prod_{i}\frac{\pi}{a_{i}}.
\end{equation}
We see a remarkable property: the product of the two integrals is
equal to 1. This property remains true for any Gaussian integrals of
commuting and anti-commuting variables. In particular,
\begin{equation}
\label{1-Z} Z^{-1}=\int {\cal D}\mu^{*}\,{\cal D}\mu\,{\rm
exp}\left[ iS_{\pm}[\mu]\right].
\end{equation}
Now the Green's functions can be represented without the
denominator:
\begin{equation}
\label{GF1} G_{nm}^{R/A}= \mp i\,\int
 {\cal D}\psi\,\,\;\;\varphi_{n}\,\varphi_{m}^{*}\,{\rm exp}\left[
iS_{\pm}[\psi]\right],
\end{equation}
where
\begin{equation}
\label{sup-act} S_{\pm}[\psi]=S_{\pm}[\varphi]+S_{\pm}[\mu]=
\pm\sum_{i,j}\psi_{i}^{\dagger}\,\left[E_{\pm}\delta_{ij}-H_{ij}
\right]\,\psi_{j}.
\end{equation}
Here we introduced the {\it super-vectors} $\psi$ and
$\psi^{\dagger}$:
\begin{equation}
\label{sup-vect}
\psi^{\dagger}=(\varphi^{*},\mu^{*}),\;\;\;\;\;\psi=\left(\begin{matrix}
\varphi\cr \mu\end{matrix} \right)
\end{equation}
and the super-measure:
\begin{equation}
\label{sup-measure} {\cal D}\psi={\cal D}\varphi^{*}\,{\cal
D}\varphi\,{\cal D}\mu^{*}\,{\cal D}\mu.
\end{equation}
The action Eq.(\ref{sup-act}) and the integration measure
Eq.(\ref{sup-measure}) are {\it super-symmetric}, i.e. the commuting
and anti-commuting variables enter in a fully symmetric way. The
super-symmetry is however broken in the {\it pre-exponent} in
Eq.(\ref{GF1}), as it depends only on the commuting variables.

Now when the problem of denominator is solved by the supersymmetry
trick, the next step is to average over the Gaussian ensemble of
$H_{ij}$. To this end we write:
$$
\mp i\sum_{ij}\psi^{\dagger}_{i}\psi_{j}H_{ij} =\frac{\mp
i}{2}\sum_{ij}\left(\psi^{\dagger}_{i}\psi_{j}H_{ij}+\psi^{\dagger}_{j}\psi_{i}H_{ji}\right)
$$
Averaging of the r.h.s. is done independently for each pair of $i,j$
using the identity:
$$
r.h.s.-\frac{1}{A_{ij}}\,|H_{ij}|^{2}=-\frac{1}{A_{ij}}\,
\left(H_{ij}\pm\frac{i}{2}A_{ij}\,\psi_{i}^{\dagger}\psi_{j}
\right)\,\left(H_{ji}\pm\frac{i}{2}A_{ij}\,\psi_{j}^{\dagger}\psi_{i}
\right)-\frac{A_{ij}}{4}\,\psi^{\dagger}_{i}\psi_{j}\psi^{\dagger}_{j}\psi_{i}.
$$
From now on for simplicity we will consider the case $\beta=2$. Then
$$
\int dH_{ij}dH_{ij}^{*}\,{\rm
exp}\left(r.h.s.-\frac{1}{A_{ij}}\,|H_{ij}|^{2} \right)=\int
d\tilde{H_{ij}}d\tilde{H_{ij}^{*}}\,{\rm exp}\left(
-\frac{1}{A_{ij}}\,|\tilde{H_{ij}}|^{2}\right)\;\;{\rm exp}\left(
-\frac{A_{ij}}{4}\,\psi^{\dagger}_{i}\psi_{j}\psi^{\dagger}_{j}\psi_{i}\right),
$$
where
$$
\tilde{H_{ij}}=H_{ij}\pm\frac{i}{2}\psi_{i}^{\dagger}\psi_{j}.
$$
The simplicity of the case $\beta=2$ is that $\tilde{H_{ij}}$
belongs to the same manifold of complex numbers as $H_{ij}$, so that
one may replace in the integral over the entire manifold
$\tilde{H_{ij}}\rightarrow H_{ij}$. Thus on the right hand side we
obtain the normalization integral for the random matrix ensemble
averaging. So we obtain for the disorder average:
\begin{equation}
\label{av} \left\langle{\rm exp}\left(\mp
i\sum_{ij}\psi^{\dagger}_{i}\psi_{j}H_{ij} \right)
\right\rangle={\rm exp}\left(-\frac{1}{4}\sum_{ij}
A_{ij}\,\psi^{\dagger}_{i}\psi_{j}\psi^{\dagger}_{j}\psi_{i}\right).
\end{equation}
Now we define the super-matrix
\begin{equation}
\label{Q-til} \tilde{Q}_{i}=\psi_{i}\otimes\psi_{i}^{\dagger}=\left(
\begin{matrix} \varphi_{i}\varphi_{i}^{*}&\varphi_{i}\mu_{i}^{*}\cr
\mu_{i}\varphi_{i}^{*}& \mu_{i}\mu_{i}^{*}\cr
\end{matrix}\right)\equiv\left(
\begin{matrix} BB&BF\cr
FB& FF\cr
\end{matrix}\right)
\end{equation}
and the super-trace:
\begin{equation}
\label{st} {\rm STr}\left(
\begin{matrix} BB&BF\cr
FB& FF\cr
\end{matrix}\right)=BB-FF.
\end{equation}
Then Eq.(\ref{av}) can be conveniently rewritten as
$$
{\rm exp}\left(-\frac{1}{4}\sum_{ij}A_{ij}\,{\rm
STr}[\tilde{Q}_{i}\tilde{Q}_{j}] \right).
$$
Finally, the averaged Green's function can be represented as
follows:
\begin{equation}
\label{GF2} \langle G_{nm}^{R/A}\rangle= \mp i\,\int
 {\cal D}\psi\,\,\;\;\varphi_{n}\,\varphi_{m}^{*}\,{\rm exp}\left[
-F[\tilde{Q}]\right],\;\;\;\;\;F[\tilde{Q}]=\mp iE_{\pm}\sum_{i}{\rm
STr}[\tilde{Q}_{i}]+\frac{1}{4}\sum_{ij}A_{ij}\,{\rm
STr}[\tilde{Q}_{i}\tilde{Q}_{j}],\;\;\;\;\tilde{Q}_{i}=\psi_{i}\otimes\psi_{i}^{\dagger}
\end{equation}
Thus we derived the {\it deterministic field theory} which is
equivalent to the Gaussian random matrix ensemble and is suitable to
compute the average Green's function. One can see that the matrix of
variances
$$
A_{ij}=\langle |H_{ij}|^{2}\rangle
$$
plays a role of the coupling constant ("coupling matrix") in the
corresponding action $F[\tilde{Q}]$.

The filed-theory representation Eq.(\ref{GF2}) can be extended to
consider the averaged product of $\langle G^{R}G^{A}\rangle $ which
is necessary to be able to compute the {\it two-point} correlation
functions. To this end, one introduce the double set of commuting
and anti-commuting variables: one for $G^{R}$ and another for
$G^{A}$ and then repeats with minor modifications all the above
steps:
\begin{equation}
\label{R-A}
 \left\langle G_{nm}^{R}\left(E+\frac{\omega}{2}\right)G^{A}_{n'm'}\left(E-\frac{\omega}{2}\right)\right\rangle= \,\int
 {\cal D}\Psi\,\,\;\;\varphi^{R}_{n}\,\varphi^{R*}_{m}\,\varphi^{A}_{n'}\,
 \varphi^{A*}_{m'}\;\;{\rm exp}\left[
-F[\bar{Q}]\right],
\end{equation}
where the action $S[\bar{Q}]$ takes the form:
\begin{equation}
\label{SS} F[\bar{Q}]=-iE\sum_{i}{\rm
STr}[\bar{Q}_{i}]-i\frac{\omega+i0}{2}\,\sum_{i}{\rm STr}[\Lambda
\bar{Q}_{i}]+\frac{1}{4}\sum_{ij}A_{ij}\,{\rm
STr}[\bar{Q}_{i}\bar{Q}_{j}],\;\;\;\;\bar{Q}_{i}=\Psi_{i}\otimes\bar{\Psi}_{i},
\end{equation}
with
\begin{equation}
\label{Psi}
\bar{\Psi}=\left(\begin{matrix}\varphi^{R*}&\mu^{R*}&-\varphi^{A*}&-\mu^{A*}\cr
\end{matrix} \right),\;\;\;\;\Psi=\left(\begin{matrix}\varphi^{R}\cr\mu^{R}\cr\varphi^{A}\cr
\mu^{A}\cr
\end{matrix} \right),\;\;\;\;\Lambda=\Lambda_{2}\otimes {\bf 1}=\left(\begin{matrix}1&0&0&0\cr
0&1&0&0\cr0&0&-1&0\cr 0&0&0&-1\cr \end{matrix} \right).
\end{equation}
There is another, in some sense {\it dual}, field-theory
representation similar to Eq.(\ref{R-A}). In contrast to
Eq.(\ref{R-A}) it involves the {\it inverse coupling matrix}
$(A^{-1})_{ij}$ rather than the variance matrix $A_{ij}$. In order
to obtain this representation one makes the {\it
Hubbard-Stratonovich transformation}:
\begin{equation}
\label{HS} {\rm exp}\left\{-\frac{1}{4}\sum_{ij}A_{ij}\,{\rm
STr}[(\Psi_{i}\otimes\bar{\Psi}_{i})(\Psi_{j}\otimes\bar{\Psi}_{j})]
\right\}=\int {\cal D}P\, {\rm
exp}\left\{-\sum_{ij}A^{-1}_{ij}\,{\rm
STr}[P_{i}P_{j}]+i\sum_{i}{\rm
STr}[(\Psi_{i}\otimes\bar{\Psi}_{i})P_{i}]\right\},
\end{equation}
where $P_{i}$ is a super-matrix field.

If one substitutes the last term in Eq.(\ref{SS}) for Eq.(\ref{HS})
the remaining integral over the $\Psi$ fields is Gaussian which can
be done using a generalization of Eqs.(\ref{int-Gr})-(\ref{int-com})
for the case where the super-matrix $K_{i}$ is not proportional to
the unity matrix (for which ${\rm STr[\ln K_{i}]}=0$):
\begin{equation}
\label{Guss-int} \int {\cal D}\Psi\;\;{\rm exp}\left\{
\bar{\Psi}_{i}\,K_{i}\,\Psi_{j}\right\}={\rm
exp}\left\{-\sum_{i}{\rm STr}[\ln K_{i}] \right\}.
\end{equation}
Now we are in a position to write down the full action of the dual
representation:
\begin{equation}
\label{dual} F[P]=\sum_{ij}A^{-1}_{ij}\,{\rm
STr}[P_{i}P_{j}]+\sum_{i}{\rm STr}[\ln(E-P_{i}+(\omega/2)\,
\Lambda)].
\end{equation}
In order to appreciate the possibilities this representation is
offering and also for further simplifications of Eq.(\ref{dual}) we
compute the inverse coupling matrix $A^{-1}_{ij}$ for the important
case of the banded random matrix ensembles where the variance matrix
$A_{ij}$ is given by Eq.(\ref{band}).

The matrix element of a matrix inverse with respect to a matrix
$A_{ij}=A(i-j)$ is given by:
\begin{equation}
\label{inV}
A^{-1}_{ij}=\int_{-\pi}^{\pi}\frac{dk}{2\pi}\,G(k)\,e^{ik(i-j)},\;\;\;\;
G(k)=\left[\sum_{m=-\infty}^{+\infty}\tilde{A}(k+2\pi\,
m)\right]^{-1},\;\;\;\;\tilde{A}(k)=\int_{-\infty}^{+\infty}
dr\,A(r)\,e^{-ik r}.
\end{equation}
Note that the summation over the reciprocal lattice vector $2\pi m$
is important.

For the case of the exponential $A(r)=e^{-|r|/B}$, we have:
$$
A(k)=\frac{2B}{1+B^{2}k^{2}},\;\;\;\;\sum_{m=-\infty}^{+\infty}\frac{2B}{1+B^{2}(k+2\pi
m)^{2}}=\frac{\sinh\left(\frac{1}{B}\right)}{\cosh\left(
\frac{1}{B}\right)-\cos k}.
$$
One can see that for large $B\gg 1$ the inverse coupling matrix
$A^{-1}_{ij}$ is extremely simple:
\begin{equation}
\label{lar-B} G(k)\approx B\,(1-\cos k),\;\;\;\;A^{-1}_{ij}\approx
\frac{B}{2}\,(2\delta_{ij}-\delta_{i,j+1}-\delta_{i,j-1}),\;\;\;\;A_{0}^{-1}=
\sum_{j}A^{-1}_{ij}\approx
\frac{1}{2B}.
\end{equation}
Eq.(\ref{lar-B}) shows that locally it is just the lattice second
derivative. However, the pre-factor in front of it is large $\sim B$
and it becomes infinite for the Wigner-Dyson ensemble. This means a
large cost of variation of $P_{i}$ in the space. Let us consider
$P_{i}=P_{0}$ being independent of $i$ to the first approximation.
Then the action becomes:
\begin{equation}
\label{P0} N^{-1}\,F[P_{0}]=A_{0}^{-1}\,{\rm STr}P_{0}^{2}+{\rm
STr}\ln(E-P_{0}).
\end{equation}
Here we also neglected a term proportional to $\omega$ which is
legitimate as long as $\omega\ll \sqrt{B}$. Minimizing the action
with respect to $P_{0}$ we obtain a saddle-point equation:
\begin{equation}
\label{spe} P_{0}(E-P_{0})=A_{0}/2.
\end{equation}
The solution to this saddle-point equation is {\it degenerate}:
\begin{equation}
\label{deg}
P_{0}=\frac{1}{2}\,\left(E+iQ\,\sqrt{2A_{0}-E^{2}}\right),
\end{equation}
where $Q$ is a super-matrix obeying the constraint
\begin{equation}
\label{q21} Q^{2}=1.
\end{equation}
Another constraint comes from the requirement of super-symmetry:
\begin{equation}
\label{const2} {\rm STr}Q=0.
\end{equation}
Indeed, a super-matrix $Q$ obeying the constraint Eq.(\ref{q21})
after the diagonalization must contain only $\pm 1$ on the diagonal.
The super-symmetry (the symmetry between commuting and
anti-commuting variables) implies that it must have the same
diagonal elements corresponding to fermionic and bosonic variables
of the given type ($R$ or $A$). But then necessarily ${\rm STr}Q=0$.

Now take into account (as the first order expansion in $\omega$) the
term in Eq.(\ref{dual})proportional to the energy difference
$\omega$ and allow for slow variations of the field $Q=Q_{i}$ in
space which do not violate the saddle-point condition
Eq.(\ref{q21}). Then we obtain plugging Eq.(\ref{deg}) into
Eq.(\ref{dual}):
\begin{equation}
\label{sigma}
F[Q]=-\frac{1}{4}\,(\pi\rho\,A_{0})^{2}\,\sum_{ij}A^{-1}_{ij}\,{\rm
STr}[Q_{i}Q_{j}]-i\frac{\pi\rho\omega}{2}\sum_{i}{\rm STr}[\Lambda
Q_{i}],
\end{equation}
where
\begin{equation}
\label{rrho} \rho=\rho_{0}(E)=\frac{\sqrt{2A_{0}-E^{2}}}{\pi
A_{0}},\;\;\;\;A_{0}=\sum_{i}A_{ij}.
\end{equation}
This is the celebrated action of the {\it nonlinear} $\sigma$ {\it
model} \cite{Efet, Mirlin-Fyod}.

For the particular case of banded random matrices with $A_{ij}$
given by Eq.(\ref{band}) we obtain:
\begin{equation}
\label{sigma-band} F[Q]=-D\sum_{i}{\rm
STr}[(Q_{i}-Q_{i+1})^{2}]-i\frac{\pi\rho\omega}{2}\sum_{i}{\rm
STr}[\Lambda Q_{i}],
\end{equation}
where $D=\frac{1}{2}B\,\left(B-\frac{E^{2}}{4}\right)\sim B^{2}$.
The continuous limit of this model is the diffusive nonlinear
$\sigma$-model:
\begin{equation}
\label{sigma-band} F[Q]=-D\int dx\;{\rm STr}[(\nabla
Q)^{2}]-i\frac{\pi\rho\omega}{2}\int dx\;{\rm STr}[\Lambda Q(x)],
\end{equation}
which was originally derived by Efetov \cite{Efet} to describe the
crossover from the diffusive dynamics to the Anderson localization
in the quasi-one dimensional multi-channel disordered wire. This
demonstrates the isomorphism of the problem of quasi-one dimensional
localization and the problem of banded random matrices
\cite{Mirlin-Fyod}.

Closing this chapter we note that the derivation of Eq.(\ref{sigma})
from the exact Eq.(\ref{dual}) requires the {\it saddle-point
approximation} Eq.(\ref{spe}). Thus Eq.(\ref{sigma}) is justified
only if the energy cost of space variations of $Q_{i}$ is high. This
happens when the variance matrix $A_{ij}$ has a form of a banded
matrix which is approximately constant at $|i-j|<B$, where $B\gg 1$.
For the Wigner-Dyson ensemble $A_{ij}=1$ and the bandwidth is
maximum possible $B\sim N$ (in particular, $A_{0}=N$). In the limit
$N\rightarrow\infty$ all spacially varying configurations of the
field $Q$ ({\it non-zero modes}) are strictly forbidden. Neglecting
them we obtain the {\it zero-mode} nonlinear sigma-model \cite{Efet}
which describes the statistics of energy levels in fully chaotic
quantum systems  of confined geometry ({\it quantum dots}):
\begin{equation}
\label{ZM} F_{WD}[Q]=-i\frac{\pi\,s}{2}\;{\rm STr}[\Lambda
Q],\;\;\;\;s=\frac{\omega}{\Delta}.
\end{equation}
\section{How to compute observables: semi-circle law from the solution to a quadratic equation}
Let us demonstrate how to compute observable quantities within the
field theory using the simplest example of the mean density of
states. It is given by
\begin{equation}
\label{dOs} \rho(E)=(-2\pi i)^{-1}\,(\langle G^{R}_{nn}(E)
\rangle-\langle G^{A}_{nn}(E) \rangle)=\frac{1}{2\pi}\int {\cal
D}\Psi\;(\varphi_{n}^{R*}\varphi_{n}^{R}+\varphi_{n}^{A*}\varphi_{n}^{A})\;e^{-F[\bar{Q}]}.
\end{equation}
One can check that the pre-exponent in Eq.(\ref{dOs}) can be
represented as
\begin{equation}
\label{pre-exp-Pi}
(\varphi_{n}^{R*}\varphi_{n}^{R}+\varphi_{n}^{A*}\varphi_{n}^{A})={\rm
STr}[\Pi\,\Psi_{n}\otimes\bar{\Psi}_{n}],
\end{equation}
where
\begin{equation}
\label{Pi} \Pi=\Pi^{R}-\Pi^{A},\;\;\;\;\;\;\;
\Pi^{R}=\left(\begin{matrix}1&0&0&0\cr0&0&0&0\cr0&0&0&0\cr0&0&0&0\cr
\end{matrix}\right),\;\;\;\;\;\Pi^{A}=\left(\begin{matrix}0&0&0&0\cr0&0&0&0\cr0&0&1&0\cr0&0&0&0\cr
\end{matrix}\right).
\end{equation}

 Now we introduce an infinitesimal {\it source field} $h_{n}$
and add to the action  Eq.(\ref{SS}) a term
$$
\delta F[\bar{Q},h]=-i\sum_{i}h_{i}\,{\rm STr}[\Pi\,\bar{Q}_{i}].
$$
One can easily check that the density of states is given by a
differentiation of the partition function with respect to the field
$h$:
\begin{equation}
\label{derr} \rho(E)=\frac{1}{2\pi
i}\,\left.\frac{\partial}{\partial h_{n}}\,\int {\cal
D}\Psi\;e^{-F[\bar{Q},h]}\right|_{h\rightarrow
0},\;\;\;\;\;F[\bar{Q},h]=F[\bar{Q}]+\delta F[\bar{Q},h].
\end{equation}
Note that the additional term in the action proportional to $h$
enters exactly like the term proportional to $\omega$, so that in
the final action of the sigma-model Eq.(\ref{sigma}) one can simply
substitute
\begin{equation}
\label{L-h}
\frac{\omega}{2}\,\Lambda\rightarrow\frac{\omega}{2}\,\Lambda+h_{n}\,\Pi.
\end{equation}
Then Eq.(\ref{derr}) results in:
\begin{equation}
\label{rrrho} \rho(E)=\rho_{0}(E)\;\frac{1}{2}\int {\cal D}Q\,\;{\rm
STr}[\Pi Q_{n}]\;e^{-F[Q]},
\end{equation}
where $\rho_{0}(E)$ is given by Eq.(\ref{rrho}) and $F[Q]$ is given
by Eq.(\ref{sigma}) at $\omega=0$. We see that the quantity
$\rho_{0}(E)$ which appear in Eq.(\ref{sigma}) from the solution of
the quadratic saddle-point equation Eq.(\ref{spe}) is not
accidentally of the form of a semi-circle as the mean density of
states is proportional to it. In the case of the Wigner-Dyson
ensemble the functional $F[Q]$ at $\omega=0$ is simply zero and the
integral in Eq.(\ref{rrrho}) is a constant independent of $E$. Thus
we conclude that the semicircle law appears in this formalism from
the  solution of a quadratic saddle-point equation.

Other statistics such as the two-point correlation functions can
also be easily computed using the formalism of the nonlinear sigma
model, however not so simply as the semi-circle law. For this one
needs a proper parametrization of the matrix $Q$ which resolves the
constraints Eqs.(\ref{q21}),(\ref{const2}).
\section{Symmetry of super-matrices $\bar{Q}$ and $Q$.}
Let us return back to the derivation of the functional
representation in terms of $\bar{Q}$. It appears \cite{KOs} that by
a change of variables:
\begin{eqnarray}
\label{ChoV} \varphi^{R/A}&=&\pm i\sqrt{\lambda_{1/2}}\,\,e^{\pm
i\varphi/2+i\Omega}\,(1-\frac{1}{2}\chi^{*}_{R/A}\chi_{R/A})\\
\nonumber \mu^{R/A}&=& \pm i \sqrt{\lambda_{1/2}}\,\,e^{\pm
i\varphi/2+i\Omega}\,\chi_{R/A},
\end{eqnarray}
where $\lambda_{1/2}\geq 0$, $0\leq \varphi\leq 2\pi$,
$0\leq\Omega\leq \pi$ and $\chi_{R/A}, \chi^{*}_{R/A}$ are the new
anti-commuting variables, one can represent
$\bar{Q}=\Psi\otimes\bar{\Psi}$ in the following form:
\begin{equation}
\label{param}
\bar{Q}=U\,\bar{\Sigma}\;U^{-1}=\left(\begin{matrix}u_{R} & 0 \cr 0
& u_{A}\cr
\end{matrix} \right)\;\left(\begin{matrix}\bar{\Sigma}_{RR} & \bar{\Sigma}_{RA} \cr
 \bar{
 \Sigma}_{AR} &
\bar{\Sigma}_{AA}\cr
\end{matrix}\right)\,\left(\begin{matrix}u^{-1}_{R} & 0 \cr 0 &
u^{-1}_{A}\cr \end{matrix}\right).
\end{equation}
The beauty of this form is that the commuting and anti-commuting
variables are separated by factorization. Namely, the outer matrices
$U,U^{-1}$ containing $2\times 2$ matrices $u_{R/A}$ and
$u^{-1}_{R/A}$
\begin{equation}
\label{U}
u_{R/A}=\left(\begin{matrix}1-\frac{1}{2}\chi_{R/A}^{*}\chi_{R/A} &
-\chi_{R/A}^{*} \cr \cr \chi_{R/A} &
1+\frac{1}{2}\chi_{R/A}^{*}\chi_{R/A}\cr
\end{matrix}\right)_{BF},\;\;\;\;\;u^{-1}_{R/A}=
\left(\begin{matrix}1-\frac{1}{2}\chi_{R/A}^{*}\chi_{R/A} &
\chi_{R/A}^{*} \cr \cr -\chi_{R/A} &
1+\frac{1}{2}\chi_{R/A}^{*}\chi_{R/A}\cr
\end{matrix}\right)_{BF}
\end{equation}
are made of the anti-commuting variables. The inner matrix
$\bar{\Sigma}$
$$
\bar{\Sigma}=\left(\begin{matrix}\bar{\Sigma}_{RR} &
\bar{\Sigma}_{RA} \cr
 \bar{\Sigma}_{AR} &
\bar{\Sigma}_{AA}\cr
\end{matrix}\right)_{RA}\equiv\left(\begin{matrix}\bar{\Sigma}_{BB} & 0 \cr
 0 &
\bar{\Sigma}_{FF}\cr
\end{matrix}\right)_{BF},
$$
which is diagonal in the FB space, contains only commuting variables
with only BB sector non-zero:
\begin{equation}
\label{Db} \bar{\Sigma}_{BB}=\left(\begin{matrix}\lambda_{1}&
\sqrt{\lambda_{1}\lambda_{2}}\,e^{i\varphi} \cr
-\sqrt{\lambda_{1}\lambda_{2}}\,e^{-i\varphi} & -\lambda_{2}\cr
\end{matrix}\right)_{RA},\;\;\;\;\bar{\Sigma}_{FF}=0.
\end{equation}
One can show that the factorized form Eq.(\ref{param}) is common to
both the field $\bar{Q}$ and the {\it dual field} $Q$, with matrices
$U,U^{-1}$ being exactly the same. However the structure of the
inner matrices $\bar{\Sigma}$ and $\Sigma$ are different. Efetov has
shown \cite{Efet} that the constraints
Eqs.(\ref{q21}),(\ref{const2}) give rise to the following structure
of $\Sigma$:
\begin{eqnarray}
\label{DbEf} \Sigma_{BB}&=&\left(\begin{matrix}\lambda&
\sqrt{\lambda^{2}-1}\,e^{i\varphi} \cr
-\sqrt{\lambda^{2}-1}\,e^{-i\varphi} & -\lambda\cr
\end{matrix}\right)_{RA},\\ \label{DbEf1} \Sigma_{FF}&=&\left(\begin{matrix}\lambda_{F}&
\sqrt{1-\lambda^{2}_{F}}\,e^{i\varphi_{F}} \cr
\sqrt{1-\lambda^{2}_{F}}\,e^{-i\varphi_{F}} & -\lambda_{F}\cr
\end{matrix}\right)_{RA},
\end{eqnarray}
where $\lambda\geq 1$, $-1\leq\lambda_{F}\leq 1$, and $\varphi,
\varphi_{F}\in(0,2\pi)$.

To make practical calculations possible we also give (without
derivation) the expressions for the Jacobians of the transformation
from original variables to the variables of the above
paramerization. They are
\begin{equation}
\label{J-Os}
J[\bar{Q}]=\frac{\pi}{4\;\lambda_{1}\lambda_{2}},
\end{equation}
and
\begin{equation}
\label{J-Ef} J[Q]= \frac{1}{8(\lambda-\lambda_{F})^{2}},
\end{equation}
for the theories with coupling matrices $A^{-1}_{ij}$ and $A_{ij}$,
respectively.

 The matrices $\bar{\Sigma}_{BB}$ of the structure Eq.(\ref{Db})
as well as the matrices $\Sigma_{BB}$ of the structure
Eq.(\ref{DbEf}) can be diagonalized by the {\it pseudo-unitary}
rotation $R$:
\begin{equation}
\label{psunit}
\bar{\Sigma}_{BB}=R\,\bar{D}\,R^{-1},\;\;\;\;\;\Sigma_{BB}=R\,D\,R^{-1},\;\;\;\;\;
R^{-1}=\Lambda_{2}\,R^{\dagger}\,\Lambda_{2},
\end{equation}
where
\begin{equation}
\label{psDia}
\bar{D}=\left(\begin{matrix}|\lambda_{1}-\lambda_{2}|\;\theta(\lambda_{1}-\lambda_{2})&0\cr
0&-|\lambda_{1}-\lambda_{2}|\;\theta(\lambda_{2}-\lambda_{1})\cr
\end{matrix} \right),\;\;\;\;\;D=\Lambda_{2}\equiv\left(
\begin{matrix}1&0\cr
0&-1\cr
\end{matrix} \right).
\end{equation}
It is clear that the rotation matrix $R$ can be multiplied by a
diagonal matrix
$$
\left(
\begin{matrix}e^{i\Phi_{R}}&0\cr
0&e^{i\Phi_{A}}\cr
\end{matrix} \right)\in U(1)\otimes U(1)
$$
without violating the condition of pseudo-unitarity and without
changing the matrix $\Sigma_{BB}$ or $\bar{\Sigma}_{BB}$. To
eliminate the redundant degrees of freedom (which lead to the
divergency of the functional integrals) the group of pseudo-unitary
matrices $U(1,1)$ should be factorized as $R\;(U(1)\otimes U(1))$,
where $R$ being a factor-group $\frac{U(1,1)}{U(1)\otimes U(1)}$.

On top of that the diagonal matrix $\bar{D}$ has a free parameter
$$
\lambda_{1}-\lambda_{2}\in {\bf R}.
$$
The complete symmetry of the manifold of matrices
$\bar{\Sigma}_{BB}$ and thus the complete symmetry  of $\bar{Q}$ is:
$$
\bar{Q}\in{\bf \frac{U(1,1)}{U(1)\otimes U(1)}\otimes} {\bf R}.
$$
In contrast to that the symmetry of matrices $\Sigma_{BB}$ is simply
$ \Sigma_{BB}\in \frac{U( 1,1)}{U(1)\otimes U(1)} $. Its counterpart
$\Sigma_{FF}$ has the symmetry $ \Sigma_{FF}\in
\frac{U(2)}{U(1)\otimes U(1)}$ as it can be diagonalized by the {\it
unitary} rotation matrix $R_{F}$. The complete symmetry of the $Q$
field in the Efetov's nonlinear sigma-model is therefore:
$$
Q\in{\bf \frac{U(1,1)}{U(1)\otimes U(1)}\otimes
\frac{U(2)}{U(1)\otimes U(1)}}.
$$
Note by passing that the number of independent variables in
$\bar{Q}$ and $Q$ is different. While both have 4 anti-commuting
variables, the number of commuting variables is $2+2=4$ for the
filed $Q$ and $2+1=3$ or the field $\bar{Q}$.

  Thus we see that the duality transformation
and the saddle-point approximation not only invert the coupling
matrix $A_{ij}$ but also change the symmetry of the {\it target
space}. Such type of duality is encountered in the string theory and
is called $T-duality$.

\section{Eigenfunction statistics}
In this section we show how to compute eigenfunction statistics
using the field theory formalism. As usual, the starting point is to
express the physical quantity of interest in terms of the Green's
functions. To this end we study the product:
\begin{equation}
\label{l-m} K_{l,m}=[G^{R}_{nn}]^{l}\,[G^{A}_{nn}]^{m}=\left(
\sum_{i}\frac{|\Psi_{i}(n)|^{2}}{E-E_{i}+i\delta}\right)^{l}\;\left(
\sum_{i'}\frac{|\Psi_{i'}(n)|^{2}}{E-E_{i'}-i\delta}\right)^{m},
\end{equation}
where we used the representation of Green's functions in terms of
exact eigenfunction $\Psi_{i}(n)$ and exact eigenvalues $E_{n}$ of a
random matrix Hamiltonian. Let us multiply Eq.(\ref{l-m}) by an
infinitesimal  $\delta^{l+m-1}$ average over realizations of the
random matrix ensemble and do the limit $\delta\rightarrow 0$. This
trick singles out only one state of the double sum, the one that is
accidentally at the energy $E$:
\begin{equation}
\label{one-state} \lim_{\delta\rightarrow
0}\delta^{l+m-1}\,K_{l,m}=\lim_{\delta\rightarrow 0}\left\langle
\sum_{i}\frac{|\Psi_{i}(n)|^{2(l+m)}\;\delta^{l+m-1}}{(E-E_{i}+i\delta)^{l}(E-E_{i}-i\delta)^{m}}.
\right\rangle
\end{equation}
The smallness of the interval $|E-E_{i}|\sim \delta$ is the reason
why the power of $\delta$ is $l+m-1$ and not $l+m$. Indeed, let the
joint probability distribution function for $\Psi_{i}(n)$ and
$E_{i}$ be $P(\Psi_{i},E_{i})$. It is a smooth function of $E_{n}$
which does not change at a scale $\delta\rightarrow 0$. Then
averaging in Eq.(\ref{one-state}) can be performed as follows:
\begin{equation}
\label{av-per} \sum_{i}\int
d\Psi_{i}dE_{i}\,P(\Psi_{i},E_{i})\,\frac{|\Psi_{i}|^{2(l+m)}}
{(E-E_{i}+i\delta)^{l}(E-E_{i}-i\delta)^{m}}\approx\sum_{i}\int
d\Psi_{i}\,P(\Psi_{i},E)\,|\Psi_{i}|^{2(l+m)}\,C_{l,m},
\end{equation}
where
\begin{equation}
\label{C-C} C_{l,m}=\int_{-\infty}^{+\infty}
\frac{dE_{i}}{(E-E_{i}+i\delta)^{l}(E-E_{i}-i\delta)^{m}}=(2\delta)^{1-(l+m)}\;2\pi
i^{m-l}\frac{(l+m-2)!}{(l-1)!(m-1)!}.
\end{equation}

 Let us define also the moment of the $|\Psi_{i}(n)|^{2}$ at an
energy $E$:
\begin{equation}
\label{mom-def} \langle |\Psi_{i}|^{2 p}
\rangle_{E}=\frac{1}{\rho(E)}\,\left\langle \sum_{i}|\Psi_{i}|^{2
p}\;\delta(E-E_{i})
\right\rangle\equiv\frac{1}{\rho(E)}\,\sum_{i}\int d\Psi_{i}\int
dE_{i}\;P(\Psi_{i},E_{i})\;|\Psi_{i}|^{2 p}\;\delta(E-E_{i}).
\end{equation}
Comparing Eqs.(\ref{mom-def}),(\ref{av-per}) we arrive at:
\begin{equation}
\label{mom-G} \langle |\Psi_{i}|^{2(l+m)}\rangle_{E} =
\frac{1}{2\pi\rho(E)}\,i^{l-m}\;\frac{(l-1)!(m-1)!}{(l+m-2)!}\;\lim_{\delta\rightarrow+0}
\left\{(2\delta)^{l+m-1}\langle
[G^{R}_{nn}(E+i\delta)]^{l}[G^{A}_{nn}(E-i\delta)]^{m}\rangle\right\}.
\end{equation}
This is the expression of the eigenfunction moments in terms of the
retarded and advanced Green's functions we were looking for. One can
see that any non-trivial moment $m+l>1$ requires a non-trivial
limiting procedure.

The next standard step is to represent the average of the Green's
functions in terms of the functional integral. It begins with the
standard representation similar to Eq.(\ref{dOs}):
\begin{equation}
\label{stand}
[G^{R}_{nn}(E+i\delta)]^{l}[G^{A}_{nn}(E-i\delta)]^{m}=\frac{i^{m-l}}{l!m!}\int{\cal
D}\Psi\;(\varphi_{n}^{R*}\varphi_{n}^{R})^{l}\,(\varphi_{n}^{A*}\varphi_{n}^{A})^{m}\;
e^{-F[\bar{Q}]}.
\end{equation}
Then the analogy with Eq.(\ref{dOs}) would suggest that we write
$\varphi_{n}^{R*}\varphi_{n}^{R}={\rm STr}[\Pi^{R}\bar{Q}]$, raise
the $h{\rm STr}[\Pi^{R}\bar{Q}]$ into the exponent with the help of
the $l$-times differentiation with respect to the background field
$h$ and then switch to the super-matrix filed $Q$ as in
Eq.(\ref{rrrho}). However, in trying to do these "standard" steps we
make a mistake. The reason is that the field $\bar{Q}_{n}$ is not
slow-varying with $n$ and the background field $h_{n}$ should also
contain fast space variations. This is what makes a difference
compared to the case  of the constant in space symmetry breaking
field $\frac{1}{2}\omega \Lambda$ in Eq.(\ref{L-h}).

One possible remedy \cite{Efet} is to single out the bi-linear
combinations of $\varphi_{n}^{*}$ and $\varphi_{n}$ which do not
contain fast space variations. We show how to do this for the
product
$\varphi_{n}^{R*}\varphi^{R}_{n}\varphi_{n}^{A*}\varphi^{A}_{n}$. As
the result of averaging over random matrix ensemble should not
depend on $n$ ({\it translational invariance on the average}) one
can do the sum over $n$ and then divide the result by $N$. Switching
to the Fourier-components $\varphi(p)$we can represent this sum as
$$
\sum_{p_{1},p_{2},q}\varphi^{R*}(p_{1})\varphi^{R}(-p_{1}+q)\varphi^{A*}(p_{2})
\varphi^{A}(-p_{2}-q)=\sum_{p_{1},p_{2},q}\varphi^{R*}(p_{1})\varphi^{R}(-p_{2}+q)
\varphi^{A*}(p_{2}) \varphi^{A}(-p_{1}-q).
$$
Two sums in the above expression is a mere re-labeling of momenta,
all what is really needed is that the sum of all momenta is zero.
However, this re-labeling becomes a non-trivial operation if one
assumes that the momentum $q$ is small. In assuming so we select a
definite {\it domain of summation} such that the corresponding
bi-linear combination of $\varphi$ is slow varying in space. Then
{\it one single sum} can be presented as
\begin{eqnarray}
\label{select-slow}
&&\sum_{p_{i}}\varphi^{R*}(p_{1})\varphi^{R}(p_{2})\varphi^{A*}(p_{3})
\varphi^{A}(p_{4})\;\delta(p_{1}+p_{2}+p_{3}+p_{4})=\\ \nonumber
&&\sum_{p_{1},p_{2},q\ll
1}\varphi^{R*}(p_{1})\varphi^{R}(-p_{1}+q)\varphi^{A*}(p_{2})
\varphi^{A}(-p_{2}-q)+\sum_{p_{1},p_{2},q\ll
1}\varphi^{R*}(p_{1})\varphi^{R}(-p_{2}+q) \varphi^{A*}(p_{2})
\varphi^{A}(-p_{1}-q)+{\rm remainder}.
\end{eqnarray}
In the first term of Eq.(\ref{select-slow}) the bi-linear
combinations $\varphi^{R*}(p_{1})\varphi^{R}(-p_{1}+q)$ and
$\varphi^{A*}(p_{2})\varphi^{A}(-p_{2}-q)$ are slow, while in the
second term slow are the combinations
$\varphi^{R*}(p_{1})\varphi^{A}(-p_{1}-q)$ and
$\varphi^{A*}(p_{2})\varphi^{R}(-p_{2}+q)$.  In the remainder we
collect all terms where there is no bi-linear slow combinations. The
meaning of the above procedure of singling out the slow bi-linear
combinations is that only such combinations lead to the divergent
functional integral in the limit when $E_{+}-E_{-}=2i\delta$ tends
to zero. The average of the remainder is not singular and can be
neglected.

Performing this procedure in Eq.(\ref{stand}) one obtains $(l+m)!$
possibilities to break the product
$(\varphi_{n}^{R*}\varphi_{n}^{R})^{l}\,(\varphi_{n}^{A*}\varphi_{n}^{A})^{m}$
into the product of slow bi-linear combinations. All of them appear
to make the same contribution to Eq.(\ref{stand}). Thus one can
consider only one such term, do all the standard manipulations with
the source fields as we explained above for the case of the mean
density of states and multiply the result by $q!$. The final result
for the simplest choice $m=1$ is:
\begin{equation}
\label{mom-fin-sigma} \langle
|\Psi_{n}|^{2k}\rangle_{E}=-\frac{k}{2}\,\lim_{\delta\rightarrow
0}\left\{(2\pi\rho\delta)^{k-1}\;\int{\cal D}Q\;({\rm
STr}[\Pi^{R}Q_{n}])^{k-1}\;{\rm
STr}[\Pi^{A}Q_{n}]\;e^{-F[Q]}\right\}.
\end{equation}
One can see that it is $k!$ times larger than the one obtained by
the "naive" manipulations with the background field. We spent some
time to go into detail of this subtlety in order to show that
sometimes "exact" manipulations with the background fields are
dangerous if the fast varying components of the fields are treated
improperly or simply omitted.

This is the result of a saddle-point approximation used in the
derivation of the nonlinear sigma-model. No such danger appear for
the dual representation which did not involve any approximation:
\begin{equation}
\label{mom-fin-dual} \langle
|\Psi_{n}|^{2k}\rangle_{E}=-\frac{1}{2\,(k-1)!}\,\lim_{\delta\rightarrow
0}\left\{(2\pi\rho\delta)^{k-1}\;\int{\cal D}\bar{Q}\;({\rm
STr}[\Pi^{R}\bar{Q}_{n}])^{k-1}\;{\rm
STr}[\Pi^{A}\bar{Q}_{n}]\;e^{-F[\bar{Q}]}\right\}.
\end{equation}
One can do one more step without specifying the functionals $F[Q]$
and $F[\bar{Q}]$ using the fact that the structure of dependence of
the $Q$ and $\bar{Q}$ fields on the anti-commuting  variables
Eqs.(\ref{param}),(\ref{U}) is the same. To this end we define
\cite{Mirlin2000}the {\it generating functions} $Y[Q]$ as the
functional integral of $e^{-F[Q]}$ done over all the super-matrices
$Q_{i}$, except  the one at a space point $n$:
\begin{equation}
\label{generating} Y[Q_{n}]=\int_{Q_{i},i\neq n} {\cal
D}Q\,e^{-F[Q]}.
\end{equation}
If the generating function is known the eigenfunction moments are
given by the integral over one single super-matrix $Q_{n}$.

One can show quite generally that this function does not depend on
the anti-commuting variables. Then the integration of the
anti-commuting variables is very simple as it involves only the
pre-exponent in Eqs.(\ref{mom-fin-sigma}),(\ref{mom-fin-dual}). As
the result of this integration the additional  factor $(k-1)$
appears in these equations. However, the main thing is to understand
how it comes that the infinitesimal factor $\delta^{k-1}$ is
compensated by the integral over the super-matrix $Q_{n}$. There is
only one scenario of for this to happen in the framework of the
nonlinear sigma-model: this is to absorb $\delta$ into the variable
$\lambda\rightarrow\delta\lambda$ which can take arbitrary large
values. For this the pre-exponent in Eq.(\ref{mom-fin-sigma}) must
be proportional to $\lambda^{k-2}$ in the limit of large $\lambda$
and also the generating function $Y(Q_{n})=Y(u)$ must be a function
of one single variable $u=2\pi\rho\delta\lambda$. One can show that
this is indeed the case:
\begin{equation}
\label{mom-fin} \langle
|\Psi_{n}|^{2k}\rangle_{E}=\frac{k(k-1)}{N}\,\int_{0}^{\infty}du\,u^{k-2}\,Y(u).
\end{equation}
This is a remarkable formula, as it implies that the distribution
function of $|\Psi|^{2}$ for any unitary ensemble $\beta=2$ is:
\begin{equation}
\label{Dfun} \left.{\cal
P}(|\Psi|^{2})=N^{-1}\,\frac{\partial^{2}}{\partial
u^{2}}\;Y(u)\right|_{u=|\Psi|^{2}}.
\end{equation}
For the dual theory Eq.(\ref{mom-fin-dual}), the generating function
$\bar{Y}[\bar{Q}_{n}]$ defined similar to Eq.(\ref{generating}) may
depend on the {\it two} variables. This is because there are not one
but two {\it non-compact} variables $\lambda_{1}$ and $\lambda_{2}$
that may take arbitrary large values. Introducing new variables
$$
s=(\lambda_{1}+\lambda_{2}),\;\;\;\;\;r=(\lambda_{1}-\lambda_{2}),
$$
one obtains:
\begin{equation}
\label{mom-fin-dual1} \langle
|\Psi_{n}|^{2k}\rangle_{E}=\frac{1}{4\pi\rho\,N}\,\frac{1}{(k-2)!}\,\int_{0}^{\infty}ds\;\int_{-\infty}^{+\infty}dr\;
\,s^{k-2}\,\bar{Y}(s,r).
\end{equation}
 Let us apply
Eq.(\ref{Dfun}) to the simplest case of the eigenfunction statistics
in the Wigner-Dyson random matrix theory. In this case the variables
of the super-matrix $Q_{i}$ are locked to their values at $i=n$.
Thus there is no integration in Eq.(\ref{generating}) whatsoever and
we obtain:
\begin{equation}
\label{Y-WD} Y[Q]={\rm exp}\left[-\pi N\rho\delta\;{\rm STr}[\Lambda
Q_{n}] \right]\rightarrow {\rm exp}(-2\pi\rho\delta\lambda\;N).
\end{equation}
Then Eq.(\ref{Dfun}) immediately gives for $\beta=2$ Wigner-Dyson
RME the Gaussian eigenfunction distribution:
\begin{equation}
\label{Dfun-WD} {\cal P}(|\Psi|^{2})= N\;e^{-N|\Psi|^{2}}.
\end{equation}
Note that the Gaussian form of the distribution function is not a
consequence of the Gaussian distribution of the entries of ${\bf H}$
but rather a consequence of the {\it central limit theorem} at any
distribution of independently fluctuating entries which variance
matrix $A_{ij}$ does not depend on $i-j$. One can show that for the
orthogonal Gaussian ensemble $\beta=1$ and for the symplectic
Gaussian ensemble $\beta=4$ it deviates from the Gaussian:
\begin{equation}
\label{Dfun-orth}
 {\cal P}(|\Psi|^{2})= \left(\frac{\beta\, N |\Psi|^{2}}{2}\right)^{\frac{\beta}{2}}\;\frac{
 e^{-\beta\,N|\Psi|^{2}/2}}{N |\Psi|^{2}\,\Gamma(\beta/2)}.
\end{equation}
This is the selebrated {\it Porter-Thomas} distribution.

The simplest non-trivial application \cite{KOs} of
Eq.(\ref{mom-fin-dual1}) for the problem that cannot be treated by
the nonlinear sigma-model is calculating the eigenfunction
distribution function for the one-dimensional Anderson model. This
model is described by the random matrix Hamiltonian
$$
 {\bf H}_{ii}=\varepsilon_{i},\;\;\;\;{\bf H}_{i,i\pm 1}=1,
$$
where $\varepsilon_{i}$ is the Gaussian random variable with the
variance $w\ll 1$ (weak disorder case). Outside the center of the
band $E=0$ the result for the eigenfunction distribution function is
amazingly simple:
\begin{equation}
\label{And-Dfun} {\cal P}(|\Psi|^{2})=\frac{V_{\rm
loc}}{N}\;\frac{e^{-V_{\rm loc}|\Psi|^{2}}}{|\Psi|^{2}},
\end{equation}
where $V_{\rm loc}$ is the localization radius. This distribution is
not normalizable and should be cut a small values of $|\Psi|^{2}$.
However, there is another way of normalizing it. This is the
requirement that $\langle |\Psi_{n}|^{2}\rangle =N^{-1}$. The first
moment of the distribution is perfectly well defined and gives the
above pre-factor.


\end{document}